\documentclass[a4paper,11pt]{article}
\usepackage{jheppub} 
\usepackage{tikz}
\usepackage{simplewick}
\usepackage{ulem}
\usepackage{bm}
\newcommand{\be}{\begin{equation}}
\newcommand{\ee}{\end{equation}}

\def \nn {\nonumber}

\def \nn {\nonumber}

\def \b {\beta}
\def \th {\theta}

\def \r {\rho}

\newcommand{\bea}{\begin{eqnarray}}
\newcommand{\eea}{\end{eqnarray}}

\newcommand{\ba}{\begin{aligned}}
\newcommand{\ea}{\end{aligned}}
\newcommand{\p}{\partial}

\title{Non-conformal Line Defect (Shell Operator) in AdS$_3$/CFT$_2$: Spinning and Higher Point Correlators}

\author[a]{Yuefeng Liu} 
\author[b,c]{and Boyang Yu}
\affiliation[a]{Institute for Advanced Study,
Kyushu University, Fukuoka 8190395, Japan}
\affiliation[b]{School of Mathematics and Maxwell Institute for Mathematical Sciences,
University of Edinburgh, Edinburgh EH9 3FD, UK}
\affiliation[c]{Center for High Energy Physics, Peking University, No.5 Yiheyuan Rd, Beijing 100871, P.
R. China}

\emailAdd{liu.yuefeng.650@m.kyushu-u.ac.jp, v1byu33@ed.ac.uk}

\abstract{Recently, a special type of non-conformal line defect, known as thin-shell operator, has played a key role in demonstrating the chaotic nature of the high energy sector in AdS$_3$/CFT$_2$. The chaotic nature was revealed concretely through a matching among the vacuum Virasoro block in holographic CFT$_2$, ETH analysis, and gravitational on-shell partition function in AdS$_3$ with nontrivial backreaction. In this work, we generalize this matching in two ways. First, we compute two-point correlator of the spinning defects, in contrast to previous scalar defect correlator, in both the microcanonical ensemble and the canonical ensemble. Holographically, these spinning defects correspond to bulk domain walls composed of dust particles with angular momentum. Using the first order formalism of gravity, it is shown that the junction condition deviates from Israel's junction condition, resulting in a discontinuous metric across the domain wall. Second, we calculate general higher point correlators involving multiple scalar defects and provide a detailed example with four defects. We see explicitly that, because line operators in CFT$_2$ are codimension one objects, the correlators depend on the order in which these nonlocal defects are inserted, unlike the Euclidean correlators of local operators. In both generalizations, we achieve a precise matching between field theory solutions, ETH analysis and gravitational on-shell actions.}

\begin{document}
\maketitle
\flushbottom

\section{Introduction}

The AdS/CFT correspondence provides a powerful duality relating gravitational theories in Anti-de Sitter (AdS) spacetime to conformal field theories (CFTs) defined on the asymptotic boundary of AdS \cite{Maldacena:1997re}. This duality has played a fundamental role in understanding quantum gravity and offers a powerful tool to compute observables in strongly coupled CFTs \cite{Witten:1998qj,Gubser:1998bc}.

Inspired by the AdS/CFT correspondence, recent developments have revealed that  low energy effective theories of gravity can capture statistical information about high energy eigenstates and interactions of their microscopic, UV completion counterparts through higher topology partition functions \cite{Saad:2019lba, Penington:2019kki, Almheiri:2019qdq, Belin:2020hea, Cotler:2021cqa,  Chandra:2022bqq}, even without explicitly specifying microscopic states in string theory or quantum gravity \cite{Strominger:1996sh, Mathur:2005zp}. In particular, the high energy sectors of the boundary holographic CFTs exhibit chaotic behavior: their spectra follow patterns predicted by the random matrix theory \cite{Altland:2021rqn, DiUbaldo:2023qli}, encoded geometrically by Euclidean wormholes. Furthermore, OPE coefficients involving heavy operators can be effectively treated as random variables drawn from Gaussian distributions \cite{Belin:2020hea, Anous:2021caj, Chen:2024hqu, Chandra:2024vhm, deBoer:2024mqg} which are also captured by gravitational wormhole solutions. The emergence of higher topological geometries encoding statistical information about boundary CFTs is intimately connected to the ensemble average and factorization problem in the  AdS/CFT correspondence \cite{Schlenker:2022dyo, Cotler:2022rud}.

These recent developments are broadly consistent with the eigenstate thermalization hypothesis (ETH) in the context of AdS/CFT \cite{Karlsson:2021duj}. ETH provides a guiding principle for understanding how high energy eigenstates in chaotic quantum systems emulate thermal ensembles \cite{Srednicki:1994mfb, Deutsch:1991msp, Rigol:2007juv,DAlessio:2015qtq, deBoer:2023vsm}. More explicitly, it 
provides a statistical description for the matrix elements of operators in CFT,
\be \langle E_i|O |E_j \rangle =g_1(\bar{E}) \delta_{ij}+  e^{-S(\bar{E})/2} g_2(\bar{E},\omega) R_{ij}\ , \quad \bar{E}=\frac{E_i+E_j}{2}\ , \quad  w=E_i-E_j\ ,  \label{eth1}  \ee 
where $|E_{i,j}\rangle$ are high energy eigenstates in the chaotic sector, $g_{1,2}$ are smooth envelope functions, and $R_{ij}$ is a complex Gaussian random variable. The expected chaotic nature of high energy sectors in AdS/CFT aligns well with various phenomena, such as the saturation of the Lyapunov exponent in the exponential growth of out-of-time-order correlators (OTOCs) \cite{Maldacena:2015waa} in holographic CFTs; heavy-heavy-light-light (HHLL) correlators in holographic CFT$_2$ at large central charge \cite{Fitzpatrick:2015zha}, which reproduce correlators in a BTZ black hole background indicating that heavy operator insertions create states indistinguishable from thermal states to low energy probes; and the spectral form factor of holographic systems \cite{Cotler:2016fpe, Saad:2018bqo, Cotler:2020ugk, Altland:2022xqx, Saad:2022kfe}, which exhibits the characteristic ramp-plateau structure indicative of chaotic dynamics. Both HHLL and OTOC correlators exemplify the importance of higher point correlation functions. In this work, we similarly investigate higher point correlators, but involving nonlocal operators.

The developments mentioned above primarily focus on local operators. Quantum field theories (QFTs), however, allow not only local operators but also extended objects defined on higher dimensional submanifolds, probing nonlocal phenomena inaccessible by local observables \cite{Andrei:2018die}. Such extended nonlocal objects are called defects. Defects have become closely tied to active research in higher form symmetries \cite{Gaiotto:2014kfa}, non-invertible symmetries \cite{Bhardwaj:2017xup, Chang:2018iay} and entanglement entropy \cite{Calabrese:2004eu, Cardy:2013nua, Bianchi:2015liz}. Although a full classification of defects remains challenging, certain subclasses, such as topological defects and conformal defects that preserve subsets of conformal symmetry, are tractable and under control \cite{Bhardwaj:2023kri, Shao:2023gho}. It is thus natural and insightful to explore how these extended defects behave in the high energy regime of AdS/CFT. While significant attention has been devoted to statistical properties of local operators, few works have addressed statistical aspects of nonlocal operators or defects in field theory \cite{Sasieta:2022ksu, Chen:2024hqu, Chandra:2024vhm}. One technical difficulty is that  computing defect partition functions or correlation functions  explicitly in defect CFTs is challenging.

Recently, significant interests have arisen in a specific type of non-conformal line defect also known as thin-shell operator in AdS$_3$/CFT$_2$ \cite{Anous:2016kss, Sasieta:2022ksu, Chen:2024hqu, Chandra:2024vhm}. These defects are constructed as a continuous limit of products of identical local operators placed along a line. They break conformal symmetries explicitly because the nontrivial configuration of the constituent local operators violates the $SL(2,\mathbb{R})$ symmetry along the defect. Due to the absence of sufficient symmetry constraints, direct field theory analysis of their correlation functions are challenging. However, the holographic dual descriptions are more transparent. For sufficiently massive defects (masses scale linearly with central charge $c$), the semiclassical bulk geometries are backreacted by a spherical domain wall composed of dust particles, describable as a perfect fluid. The resulting backreacted solutions can be straightforwardly obtained using Israel's junction condition, enabling the construction of stable wormhole geometries to probe statistical properties of extended defects in AdS/CFT. Holographically, thin-shell operators have been used to model black hole microstates, study black hole entropy \cite{Balasubramanian:2022gmo, Balasubramanian:2022lnw, Climent:2024trz, Geng:2024jmm} and facilitate the exploration of Hilbert space factorization in eternal black holes \cite{Balasubramanian:2024yxk, Li:2024nft}.

Given their utility in gravity, it is natural to ask whether these non-conformal defects could be studied analytically  from the field theory perspective. Building upon earlier work \cite{Anous:2016kss}, where such defects were used to generate collapsed   black holes states, \cite{Chen:2024hqu} addressed this question directly by studying the correlators of line defects
using the monodromy method and the vacuum Virasoro block approximation. It was shown that the nonlocal structure of the line defects simplifies the problem significantly compared to local operator insertions. Specifically, the Fuchsian differential equation becomes exactly solvable in the continuous limit, yielding a non-perturbative solution for the vacuum Virasoro block that dominates in the large $c$ limit. This allows a direct comparison with gravitational on-shell action  where a precise matching has been established to the leading order in the large $c$ limit.
Furthermore, \cite{Chandra:2024vhm} also studied the statistical properties of the  line defect and established a triangular equivalence among two dimensional Liouville field theory, compact holographic CFT$_2$, and AdS$_3$ gravity. 
These developments not only enrich the study of holographic on-shell actions but also provide genuine field theoretic calculations supporting the stastistical behaviors of heavy operators in the high energy sector of AdS/CFT. Additionally, from a defect CFT viewpoint, these  computations offer robust analytic control   for non-conformal line defects.

\begin{figure}
    \centering
    \includegraphics[width=0.9\linewidth]{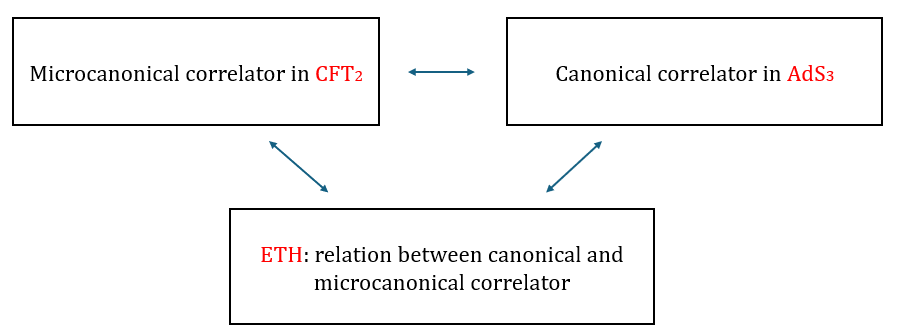}
    \caption{{The figure shows the type of  line defect correlators that will be studied in our work. Correlators computed from three different perspectives satisfy  the triangular equivalence, where  ETH ansatz serves as a bridge that effectively connects the microcanonical correlator computed in CFT with the canonical correlator in gravity.}}
    \label{fig:enter-label}
\end{figure}

Since these line defects are heavy enough, their backreaction on the geometry cannot be ignored. These dynamical effects, which contain information beyond symmetry constraints, can be successfully captured by the vacuum Virasoro conformal block \cite{Anous:2016kss}. Their heaviness  introduces technical subtleties in matching  correlators computed in field theory to   gravitational on-shell actions, which becomes apparent when comparing correlators computed in the microcanonical and canonical ensembles. More concretely, for light probe operator $\mathcal{O}$, the black hole state  with inverse temperature $\beta$ can be well approximated  by a heavy primary state $|E\rangle$ in CFT \cite{Hartman:2013mia, Asplund:2014coa, Fitzpatrick:2015zha}. We have 
\be\label{ETH-obs}
\langle E| \mathcal{O}| E\rangle \approx\langle\mathcal{O}\rangle_{\beta=\b_E }\ ,
\ee
where $\beta_E$ is the effective temperature set by the energy $E$ of the primary state. {When the observables   are heavy enough so that probe corrections cannot be ignored, \eqref{ETH-obs} can still be established at the leading order in the large $c$ expansion, but with a corrected
 matching temperature $\beta^*$ which deviates from $\beta_E$
\cite{Chen:2024hqu}. In general, $\beta^*$ is a complicated function determined by both $E$ and the observable $\mathcal{O}$.} This phenomenon has also been observed in AdS/CFT for other operator configurations \cite{Lin:2016dxa, Faulkner:2017hll}. {In short, due to the heaviness of defects,   ETH in \eqref{ETH-obs} differs from its usual form for light probes, providing a nontrivial extension of ETH in holographic CFTs. }

{In this work, we study the correlators of line defects within the triangular framework illustrated in figure \ref{fig:enter-label} by
extending \cite{Chen:2024hqu} in two ways.} We first generalize \cite{Chen:2024hqu} by considering defect operators with nonzero spin.  Although scalar operators have been studied extensively in CFT, adding spinning degrees of freedom can reveal fruitful physics, such as the  entanglement entropy with gravitational anomalies \cite{Castro:2014tta} and the discontinuity of holographic complexity \cite{Chen:2023tpi}, which go beyond non-spinning analysis. Spinning defects are also studied in \cite{Lauria:2018klo, Kobayashi:2018okw}. Here, we apply our triangle framework to
study the correlators of spinning  defects. On the CFT side, we derive the monodromy equations for the Virasoro vacuum blocks, which are shown to factorize into holomorphic and anti-holomorphic pieces. This factorization implies that the thermal correlators of spinning defects also split into independent holomorphic and anti-holomorphic parts. On the  gravity side, we model the defects as collection of dust particles in rotating BTZ black hole backgrounds. We construct the backreacted saddle geometries by gluing two rotating BTZ patches along the  worldvolume of the dust. For non-spinning case, Israel’s junction condition can apply directly. When the particles carry spin, however, the spin current couples to the bulk gravity via a torsion term and Einstein equations suggest that the induced metric on the worldvolume cannot be continuous \cite{Watanabe:2004nt}. Following \cite{Giacomini:2006um}, we instead require only the continuity of the volume form.  With the relaxed junction condition, we can determine the backreacted geometries and obtain the thermal correlators by evaluating the Euclidean on-shell actions. Finally, we demonstrate precise agreement among the  monodromy methods in field theory, ETH analysis, and the gravitational on-shell action.

We then study higher point correlators across the three parts with focus on non-spinning defects. On the bulk side, the extensions are straightforward since the backreacted geometries are still  BTZ black holes away from the domain walls.
Using Israel’s junction condition, we can glue successive BTZ patches across multiple domain walls to build a semiclassical geometry with a single boundary. On the field theory side, the computations are more involved. We must solve increasingly intricate monodromy conditions beyond the two-defect case. We demonstrate this explicitly for four defect insertions, from which the general structure of higher-point  correlators  and their matching emerges. Since OTOCs and other diagnostic quantities reside in higher point correlators of local operators, we anticipate that our analysis of higher point defect correlators will likewise uncover rich chaotic and thermalization physics.

The paper is organized as follows. In Section 2, we review \cite{Chen:2024hqu} for the computation of two-point functions of non-spinning defect operators and illustrate how the three elements in the triangle match with each other. In Section 3, we generalize both ETH analysis and CFT computations to study the correlators of spinning defect operators, where a precise agreement between the two methods can be shown. In section 4, we study the thermal correlator for both non-spinning and spinning defects at arbitrary temperature $\beta$ and angular potential $\theta$ from holographic perspective. In particular, we determine the backreaction of spinning dust particles to the rotating BTZ black hole using the relaxed junction condition.  We then show that the thermal correlator, obtained by evaluating the on-shell action of the saddle geometry, agrees with the results obtained from the ETH and field theory analysis.
In Section 5, we extend our methods to higher point defect correlators. We first present explicit calculations of  four point functions as an concrete example and then extracting the general computational patterns for $2n$ point functions. Section 6 outlines several future directions based on the results we have obtained. The appendix provides  the derivation of junction condition in presence of spinning defects using the first order formalism.

\section{Review: Scalar Defect Correlators}

In this section, we follow \cite{Chen:2024hqu} and review the three corners of the triangle in figure \ref{fig:enter-label}
by examining the two-point correlators of line defects with vanishing spin. Throughout this paper, we focus only on the leading order result in the large $N$ expansion. In the context of AdS$_3$/CFT$_2$, $N$ is related to the central charge $c$ and Newton's constant $G$ by \cite{Brown:1986nw}
\be
N\sim c=\frac{3}{2G}\ ,
\ee
with the AdS radius set to be unit.
We are interested in both the microcanonical correlator and renormalized canonical correlator defined as follows
\be\ba\label{correlator}
G_{E_H}(t)\equiv\langle E_H|D^\dagger(t)D(0)|E_H\rangle\ ,\quad 
G_\beta(t)\equiv Z_\beta^{-1}\text{Tr}\left(e^{-\beta E_H}G_{E_H}(t)\right)\ ,
\ea
\ee
where $|E_H\rangle=O_H(t=-\infty)|0\rangle$ is a high energy eigenstate created by a local primary operator $O_H$ with conformal dimension $h_H=\bar{h}_H=\frac{1}{2}(E_H+\frac{c}{12})$,
and
$Z_\beta=\text{Tr}(e^{-\beta H})$ is the thermal partition function. We also require the state to be heavy enough so that $E_H>\frac{c}{12}$ is in the Cardy regime \cite{Cardy:1986ie}. 
The line defect $D(t)$ in CFT$_2$ is defined as the product of infinite identical local operators located on the constant time slice of cylinder,
\be\label{def:D}
D(t)\equiv\prod_{k=1}^n\Psi\left(t,\frac{2\pi(k-1)}{n}\right)\ ,
\ee
with $\Psi$ being a primary operator with conformal dimension $h_\Psi$. We set $h_\Psi\sim 1$ and $n\sim N$ so that the total dimension satisfies
\be
h_D\equiv nh_\Psi\sim N\sim c\ .
\ee
By computing \eqref{correlator} in both field theory and gravity theory, we demonstrate that the defect $D$ admits a statistical description suggested by the ETH ansatz \eqref{eth1} with $g_1=0$. 
This indicates that AdS$_3$/CFT$_2$ holographic system displays the chaotic feature encoded by $D(t)$. Our analysis adheres to the bottom-up philosophy of AdS/CFT without invoking any particular field theory, while there exists
an interesting explicit construction of the  defect and corresponding matching using
Liouville CFT \cite{Chandra:2024vhm}.

\subsection{ETH analysis}
We assume the line defect $D$ fits into the ansatz \eqref{eth1} with $g_1(\bar E)=0$ so that its matrix element is given by
\be 
D_{ij}=e^{-S(\bar{E})/2} g_2(\bar{E},\omega) R_{ij}=e^{-f_{h_D}(E_i,E_j)/2} R_{ij}\ . \label{ETH} \ee 
Plugging this into the microcanonical correlator $G_{E_H}(t)$ given by \eqref{correlator}, 
we get \footnote{We have replaced the discrete summation over energy eigenstate by an integral. More precisely, $\sum_E=\int dEe^{S(E)}$ with $e^{S(E)}$ being the energy density in an average sense.}
\begin{align} \label{miccol}
G_{E_H}(t)= \int dE e^{S(E)+t(E_H-E)-f_{h_D}(E_H,E)}\  .
\end{align}
We also assume that the contribution from high energy states dominates the integral so that $S(E)$ can be approximated by the Cardy entropy for states with zero spin \cite{Cardy:1986ie}
\be\label{Cardy0}
S(E)=2\pi\sqrt{\frac{c}{3}E}\ ,\quad E>\frac{c}{12}\ .
\ee
Using the saddle point approximation, \eqref{miccol} becomes
\be\label{saddle:GEh}
\log G_{E_H}(t) = S(E)+t (E_H-E)-f_{h_D}(E_H,E)\  ,
\ee
where the  value of $E=E(E_H,t)$
is determined by solving the following saddle equation
\be 
\p_{E} S(E)-t -\p_{E} f_{h_D}(E_H,E) =0\ . \label{microcod} 
\ee 
Using \eqref{microcod},  the time derivative of the correlator is simply given by the explicit time dependence in the exponent of \eqref{miccol},
\be \frac{d}{dt} \log{G_{E_H}(t)}= E_H-E(E_H,t)\ . \label{micdetau} \ee 
Similarly, the canonical correlator $G_{\beta}(t)$ can be obtained from the ETH ansatz as
\begin{align}
    {Z}_{\beta} G_{\beta}(t)&= \int dE_He^{-\beta E_H+S(E_H)}  G_{E_H}(t) = e^{S(E_H)+S(E)-tE-(\beta-t)E_H -f_{h_D}(E_H,E) }\  , \label{canotau}
\end{align}
where $E_H(\beta,t)$ in the last equality are determined by
\be  \p_{E_H} S(E_H) -\beta+ t- \p_{E_H} f_{h_D}(E_H,E)  =0  \label{cons1}\  .\ee 
The time derivative of the thermal correlator is then given by
\be \frac{d}{dt} \log { {Z}_{\beta} G_{\beta}(t)} =E_H-E\ .\label{pataucan}  \ee 
Above are results for two-point correlators of non-conformal line defects assuming the validity of ETH ansatz. As we are going to review, both \eqref{miccol} and \eqref{canotau} can be reproduced from explicit  calculations in  field theory and gravity, while the matching among them determines the envelop function $f_{h_D}$.

\subsection{Vacuum Virasoro block}
On the field theory side,
the vacuum Virasoro block of the microcanonical correlator $G_{E_H}(t)$, which is expected to give the dominant contribution, can be computed using the monodromy method \cite{Chen:2024hqu}. 
By standard exponential map $z=e^{t+i\phi}$, $G_{E_H}(t)$ is mapped to the defect correlator on the plane as 
\be  G_{E_H}(t)=\langle E_H|D^\dagger(t_2)D(t_1)|E_H\rangle=e^{2h_D(t_1+t_2)}\langle O_H(\infty) D^\dagger(r_2) D(r_1) O_H(0)\rangle\ , \ee 
with $t=t_2-t_1,r_i=e^{t_i}$. According to \cite{Zamolodchikov:1987avt}, the vacuum Virasoro block should take an exponential form in the large $c $ limit, thus we have
\be\label{GEh-F} G_{E_H}(t) \approx e^{2h_D(t_1+t_2)} e^{-\frac{c}{3}\mathcal{F}}\ .\ee 
Defining $K=\frac{\p\mathcal{F}}{\p r_1}$ and using \eqref{GEh-F}, the time derivative of $\log G_{E_H}(t)$ can be expressed as
\be\label{dt:GEh}
\frac{d}{dt}\log G_{E_H}(t) \approx 2h_D-\frac{c}{3}Kr_1\ .
\ee
The parameter $K$ can be determined by solving the monodromy equation. We leave the detailed calculations to next section when we study the correlators of spinning defects. The results for scalar correlator here can be viewed as its holomorphic sector. 
Let us now summarize the results below. If we define
\be
\rho_1=\frac{\sqrt{1-24h_H/c}}{2}\ ,\quad \rho_2=\frac{\sqrt{1-4(6h_H/c-6h_D/c+Kr_1)}}{2}\ ,
\ee
then \eqref{dt:GEh} can be rewritten as
\be\label{dt:mono}
\p_t\log G_{E_H}(t)=\frac{c(\rho_2^2-\rho_1^2)}{3}\ ,\quad E_H=-\frac{c\rho_1^2}{3}\ .
\ee
The monodromy equation for $\rho_2$ is given by
\be\label{mono-eq}
e^{2\rho_2t}=\frac{(\rho_2+\rho_1-6h_D/c)(\rho_2-\rho_1-6h_D/c)}{(\rho_2-\rho_1+6h_D/c)(\rho_2+\rho_1+6h_D/c)}\ ,
\ee
which is a transcendental equation for $\rho_2$ and can be solved either numerically or perturbative in various limits.

\subsection{Gravitational action}

\begin{figure}
    \centering
    \includegraphics[width=0.6\textwidth]{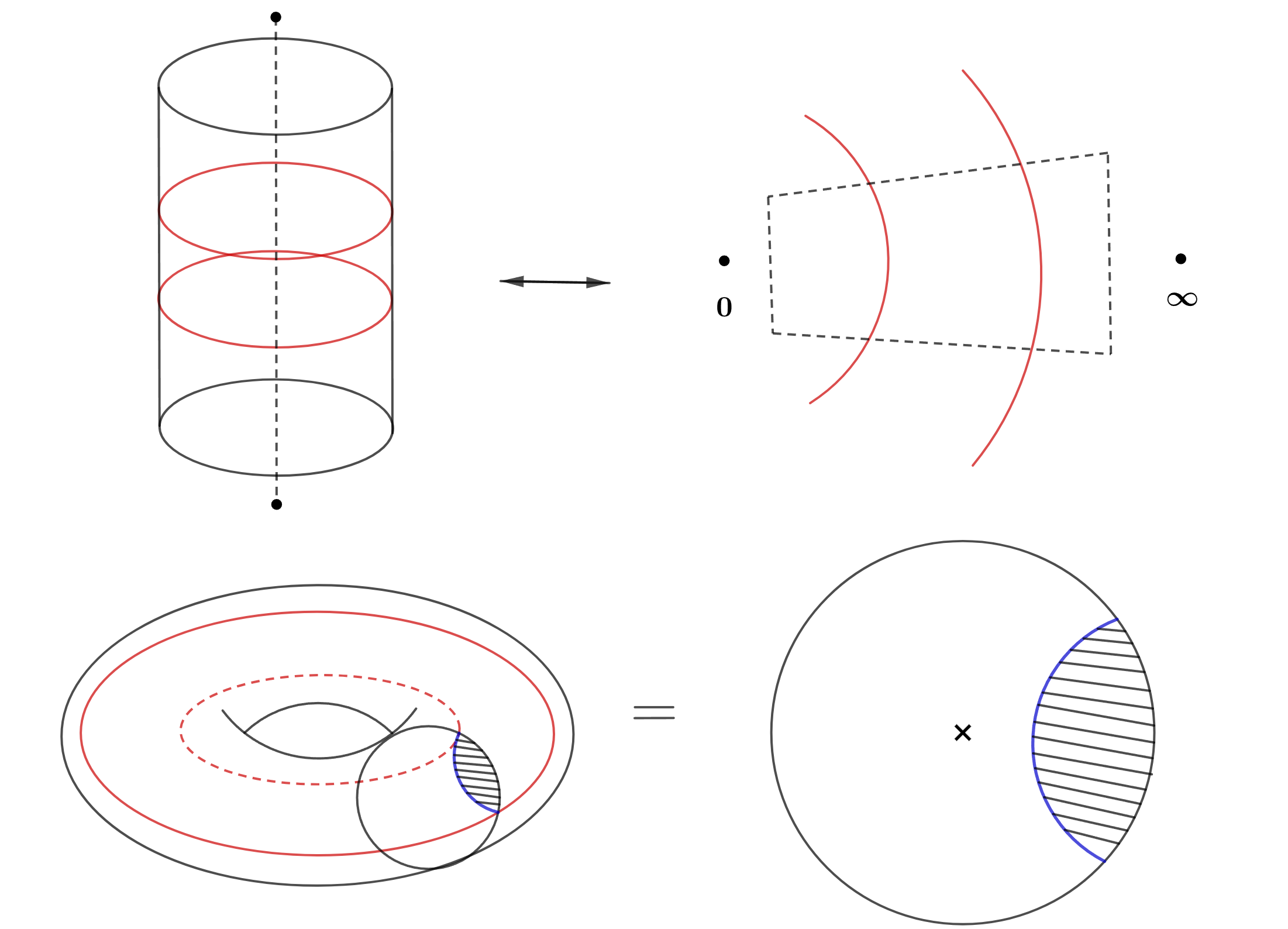}
    \caption{The AdS/CFT setup with two line defects (red). The two figures above show the microcanonical CFT$_2$ correlator $\langle E_H|D^{\dagger} D| E_H \rangle$ and its associated monodromy contour related to vacuum Virasoro block. The two figures below show the AdS$_3$ canonical correlator $\langle D^{\dagger} D\rangle_{\b}$ backreacted with dust (blue), together with its two dimensional schematic.}
    \label{fig:adscft1}
\end{figure}

On the gravity side, we focus on the holographic computation of the thermal correlator $G_\beta(t)$ defined on the torus for a compact CFT$_2$. The dual geometry of the torus could be either a BTZ black hole or thermal AdS. We set the temperature $\beta^{-1}$  high enough so that the black hole sector dominates. Since $h_D\sim c\gg1$, the line defect $D$ is holographically represented by a collection of dust particles smeared evenly over the spatial circle. To compute the thermal correlator, we need to study the geometry with the dust propagating in the bulk. The dust forms a domain wall
which starts and ends at the boundary insertion positions of $D$ and $D^\dagger$, see figure \ref{fig:adscft1}. Moreover, 
the dust is heavy enough so the nontrivial backreaction has to be taken into account even at the leading order of $G^{-1}$ expansion.

The Euclidean action of Einstein gravity with a domain wall in AdS$_3$ is given by
\be  \label{grapre}
I=-\frac{1}{16\pi G}\int\sqrt{g}(R+2)+\int_{\mathcal W}\sqrt{h} \sigma \ ,
\ee
where $\mathcal W$ denotes the worldvolume of the dust, $h$ is the induced metric and $\sigma$ is the density. The original background is chosen to be a BTZ black hole with mass $M=\frac{4\pi^2}{\beta^2}$. The backreacted geometry is divided into two BTZ geometries with masses $M_+$ and $M_-$ glued together via the Israel's junction condition
\be\label{jun1}
[g_{ab}]=0\ ,\quad [K_{ab}-g_{ab}K]=-T_{ab}\ ,
\ee
where $K_{ab}$ is the extrinsic curvature of $\mathcal{W}$, and we use $[X]$ to denote the difference of quantity $X$ evaluated on two sides of the shell, i.e. $[X]\equiv X_--X_+$. As has been shown in \cite{Sasieta:2022ksu,Chen:2024hqu}, the saddle point solutions of the backreacted geometry are solved by the following equations,
\be \label{glocond1} \frac{r_-}{r_*}=\sin\frac{r_-(\beta-t)}{2}\  , \quad  \frac{r_+}{r_*}=\sin\frac{r_+t}{2} \  , \ee
where $r_\pm=\sqrt{M_\pm}$, and $r_*$ is the radius of turning point of the domain wall \footnote{We have rescaled the mass of the dust compared to \cite{Chen:2024hqu} with $m_{\text{here}}=4Gm_{\text{there}}$.}
\be r_*^2=r_{\pm}^2+\big( \frac{M_{\pm}-M_{\mp}}{2m}-\frac{m}{2} \big)^2\ . \ee 
The regularized gravitational on shell action is computed to be
\be\label{grav-act}
 \log [Z(\beta)G_{\beta}]=-I_{\text{reg}}=\frac{tM_++(\beta-t)M_-}{8G}+\frac{m}{2G}\log\frac{r_*}{2}\ , 
 \ee 
where $r_+=r_+(\beta,m,t)$ and $r_-=r_-(\beta,m,t)$ are obtained by solving  \eqref{glocond1}. 
By taking time  derivative of \eqref{grav-act} and using \eqref{glocond1}, it can be shown that
\be \frac{d}{dt} \log[{{Z}_{\beta} G_{\beta}(t)}]=\frac{M_- -M_+}{8G}\ . \label{2grat}  \ee 
\subsection{Matching}
Now we summarize the relations of correlators computed in the three ways  described above.
\paragraph{CFT  $\Leftrightarrow$ ETH:} Using the fact that $E_H=-\frac{c\rho_1^2}{3}$, it is easy to see that \eqref{micdetau} agrees with \eqref{dt:mono} provided 
\be \label{iden1}
E=-\frac{c\rho_2^2}{3}\ .
\ee 
Such identification can be realized if we choose the envelop function to be 
\be\label{envelop}
f_{h_D}(E_H,E)=2\sqrt{\frac{cE}{3}}\arcsin\sqrt{\frac{E}{E_*}}+2\sqrt{\frac{cE_H}{3}}\arcsin\sqrt{\frac{E_H}{E_*}}-2h_D\log\frac{E_*}{2}\ ,
\ee
where
\be
E_*=\frac{3h_D^2}{c}+\frac{c(E-E_H)^2}{48h_D^2}+\frac{E+E_H}{2}\ .
\ee
With such choice, it can be proved that the saddle point equation \eqref{microcod} for $E$  is precisely the monodromy equation \eqref{mono-eq} for $-\frac{c\rho_2^2}{3}$. Therefore, \eqref{iden1} holds and the microcanonical correlator computed by ETH ansatz matches with the vacuum Virasoro block obtained by the monodromy method in holographic CFT$_2$.
\paragraph{ETH $\Leftrightarrow$ Gravity:} Remarkably, with the same choice \eqref{envelop} and the holographic dictionary
\be
c=\frac{3}{2G}\ ,\quad h_D=\frac{m}{8G}\ ,
\ee
the saddle point equation \eqref{cons1} can be shown to agree precisely with \eqref{glocond1} under the identification $E_H=\frac{M_-}{8G},E=\frac{M_+}{8G}$. By virtue of  \eqref{pataucan} and \eqref{2grat}, we also find an agreement between ETH ansatz and gravitational calculation.
\paragraph{Gravity  $\Leftrightarrow$ CFT:} In the above, we have already used AdS/CFT dictionaries to match the results. Thus it is clear that the microcanonical correlator \eqref{miccol} and canonical correlator \eqref{canotau} are related via a Laplace transformation.

In conclusion, the high energy sectors of AdS$_3$ and CFT$_2$ provide consistent realizations of ETH result and match with each other via the AdS/CFT correspondence.

\section{Spinning Defect Correlators: ETH and CFT$_2$}

Having reviewed the computations of correlators for scalar defects in both field theory and gravity, as well as their relations to the ETH ansatz,  we are ready to generalize these results to defects with nonzero spin. More precisely, we are interested in the following microcanonical and canonical correlators
\be
\ba\label{def:G-spin}
G_{E_L,E_R}(t)&\equiv\langle E_L,E_R|D^\dagger(t_1)D(t_2)|E_L,E_R\rangle\ ,\\
G_{\beta_L,\beta_R}(t)&\equiv Z_{\beta_L,\beta_R}^{-1}\text{Tr}\left(e^{-\beta_LE_L-\beta_RE_R}G_{E_L,E_R}(t)\right)\ ,
\ea
\ee
where $D(t)$ is given by \eqref{def:D}, and we allow the local primary $\Psi$ to have different holomorphic and anti-holomorphic conformal dimensions $(h_\Psi,\bar h_\Psi)$. The state $|E_L,E_R\rangle$ is created by the primary operator with energies $E_L=h+\frac{c}{24},E_R=\bar h+\frac{c}{24}$. Similarly, for the canonical correlator, we also allow the BTZ black hole background to have different left and right moving temperatures.

\subsection{ETH analysis}
We make the ansatz on the matrix element of spinning defect $D$ that generalizes \eqref{ETH} to include states with different left and right moving energies,
\be\label{ETH-spin}
D_{ij;\bar i\bar j}\equiv\langle E_i,\bar E_i|D|E_j,\bar E_j\rangle=e^{-\tilde f(E_i,\bar E_i,E_j,\bar E_j)/2}R_{ij;\bar i\bar j}\ .
\ee
With above ansatz, the microcanonical correlator is computed to be
\be\label{GELER}
G_{E_L,E_R}(t)=\int dE'_LdE'_Re^{\tilde S(E'_L, E'_R)+t(E_L+E_R-E'_L-E'_R)-\tilde f(E_L,E_R,E'_L,E'_R)}\ .
\ee
We focus on the contribution from high energy states so that $\tilde S$ is given by the Cardy entropy for states with nonzero spin \cite{Hartman:2014oaa},
\be
\tilde S(E'_L,E'_R)=2\pi\sqrt{\frac{c}{6}E'_L}+2\pi\sqrt{\frac{c}{6}E'_R}=S\left(\frac{E'_L}{2}\right)+S\left(\frac{E'_R}{2}\right)\ ,
\ee
with $S(E)$ being given by \eqref{Cardy0}. In the large $c$ limit, we use the saddle point approximation and $\log G_{E_L,E_R}(t)$ is given by the saddle value of the exponent of the integrand in \eqref{GELER} with saddle point equations for $E_L',E'_R$ being
\be
\p_{E'_a}(\tilde S-\tilde f)-t=0\ ,\quad a=L,R\ .
\ee
The time derivative of the correlator is then given by
\be
\frac{d}{dt}\log G_{E_L,E_R}(t)=E_L+E_R-E_L'-E_R'\ .
\ee
For the canonical correlator $G_{\beta_L,\beta_R}(t)$, combining the ETH ansatz \eqref{ETH-spin} with saddle point approximation gives
\be
\ba
Z_{\beta_L,\beta_R}G_{\beta_L,\beta_R}(t)=e^{-\mathcal{I}}\ ,
\ea
\ee
with
\be
\mathcal{I}=\tilde f-\sum_{a=L,R}\left[S\left(\frac{E_a}{2}\right)+S\left(\frac{E_a'}{2}\right)-(\beta_a-t)E_a-tE_a'\right]\ .
\ee
The energies $E_a,E'_a$ are solved by the saddle point equations
\be
\p_{E_a}\mathcal{I}=\p_{E'_a}\mathcal{I}=0\ ,\quad a=L,R\ .
\ee
Similar to the non-spinning case, we will show that correlators derived from the ETH ansatz \eqref{ETH-spin} match those computed from both field theory and holographic sides in the following.

\subsection{Vacuum Virasoro block}

We first perform the exponential map $z=e^{t+i\phi}$, and
the correlator \eqref{GELER} on cylinder can be transformed into the correlator on  plane \footnote{Here we assume that these operators have integer spin. There could be a time independent phase factor under the conformal transformation for operators with non-integer spin. However, we focus on derivative of correlators in this paper, so the phase factor would not change our conclusion and our result can be easily generalized to non-integer spin case.}
\be\label{cyl-plane}
G_{E_L,E_R}(t)=e^{(h_D+\bar h_D)(t_1+t_2)}G^{\text{plane}}_{ E_L,E_R}(t)\ ,\quad t=t_2-t_1\ ,
\ee
where $\bar h_D=n\bar h_\Psi$ and
\be
G^{\text{plane}}_{E_L,\bar E_R}(t)=
\langle E_L,E_R| D^\dagger(r_2)D(r_1)| E_L,E_R\rangle_{\text{plane}}\ ,
\ee
with $r_i=e^{t_i}$.
The contribution from the vacuum Virasoro block to the correlator on the plane can be computed in a similar way as \cite{Chen:2024hqu}. The only modification is to treat holomorphic and anti-holomorphic parts independently. To proceed, we
begin with the following Fuchsian equations
\be\label{diff-eq}
V''(z)+T(z)V(z)=0\ ,\quad \bar V(\bar z)''+\bar T(\bar z)\bar V(\bar z)=0\ ,
\ee
where the holomorphic stress tensor $T(z)$ is given by
\be\label{T}
T(z)=\frac{6h/c}{z^2}-\frac{c_H}{z}+\sum_{a=1}^2\sum_{k=1}^n\frac{6h_\Psi/c}{(z-z_a^{(k)})^2}-\frac{c_{\Psi,a}^k}{z-z_a^{(k)}}\ ,\quad z_a^{(k)}=r_ae^{\frac{2\pi i(k-1)}{n}}\ .
\ee
 The anti-holomorphic stress tensor $\bar T(\bar z)$ has similar expression to \eqref{T} with all quantities replaced by the anti-holomorphic counterparts. The accessory parameters $c_H$ and $c_{\Psi,a}^k$ above are determined by the boundary and monodromy conditions which we speficy below.
In the continuous limit $n\to\infty$, the infinite sum in \eqref{T}, which we denote as $T_\Psi$, becomes an integral
\be\label{Tpsi}
T_\Psi\equiv\sum_{a=1}^2\int^{2\pi}_0\frac{d\theta}{2\pi}\left[\frac{6h_D/c}{(z-r_ae^{i\theta})^2}-\frac{c_{\Psi,a}(\theta)}{z-r_ae^{i\theta}}\right]\ .
\ee
 Due to the rotational symmetry, the residue of $T(z)$ in cylinder coordinates should be independent of $\theta$. Consequently, the angular dependence of $c_{\Psi,a}$ is fixed  to be $c_{\Psi,a}=K_ae^{-i\theta}$ for some constants $K_a$. Consequently, 
the integral \eqref{Tpsi} can be computed directly and  we get
\be\ba
T_\Psi=&\frac{(6h_D/c-K_1{r_1})\Theta(|z|-{r_1})+(6h_D/c-K_2{r_2})\Theta(|z|-{r_2})}{z^2}
\\&-\frac{6{h_D}/c [{r_1}\delta(|z|-{r_1})+{r_2}\delta(|z|-{r_2})]}{z^2}\ .
\ea\ee
Since there is a local operator with holomorphic dimension $h$ inserted at infinity, the asymptotic behavior of $T(z)$ should be $T\sim\frac{6h/c}{z^2}$. This yields  following constraints on the accessory parameters
\be\label{acc1}
c_H=0\ ,\quad   \frac{12h_D}{c}=K_1{r_1}+K_2{r_2}\ .  
\ee
Similarly, for the anti-holomorphic part, we  have
\be\label{acc2}
\bar c_H=0,\quad \frac{12\bar h_D}{c}=\bar K_1{r_1}+\bar K_2{r_2}\ .
\ee
Finally, the holomorphic stress tensor in continuous limit is given by
\be\label{stress}
T=\frac{6h}{cz^2}-\frac{6{h_D}-cK_2{r_2}}{cz^2}\Theta((|z|-r_1)(r_2-|z|))-\frac{6{h_D} [{r_1}\delta(|z|-{r_1})+r_2\delta(|z|-{r_2})]}{cz^2}\ . 
\ee
In the large $c$ limit, the correlator $G_{E_L,E_R}^{\text{plane}}$ can be approximated as
\be\label{Gplane}
G_{E_L,E_R}^{\text{plane}}\approx e^{-\frac{c}{6}(\mathcal{F}+\bar{\mathcal{F}})}\ ,
\ee
with $\mathcal{F}$ and $\bar{\mathcal{F}}$ satisfying
\be\label{timed}
\frac{\p \mathcal{F}}{\p t_i}=K_ir_i\ ,\quad \frac{\p\bar{\mathcal{F}}}{\p t_i}=\bar K_ir_i\ .
\ee
To determine parameters $K_i$ and $\bar K_i$, we need to impose appropriate
monodromy conditions. The solution to the  equation \eqref{diff-eq} with $T(z)$ given by \eqref{stress} can be written as
\be
\ba\label{sol-Fuchsian}
V(z)=\left\{\begin{array}{l}
 J_1V_1    ,\quad 0\leq|z|< {r_1} \\
    V_2  ,\quad {r_1}\leq |z|\leq {r_2}\\
    J_3V_3   ,\quad |z|\geq {r_2}
\end{array}\right.\ ,
\ea
\ee
where $V_i$ are two linearly independent solutions in each region and given by
\be
V_i=(z^{1/2-\rho_i},z^{1/2+\rho_i})^T\ ,
\ee
with
\be\label{rhos}
\rho_1=\rho_3=\frac{\sqrt{1-24h/c}}{2}\ ,\quad \rho_2=\frac{\sqrt{1-4(6h/c-6h_D/c+K_2r_2)}}{2}\ .
\ee
Plugging \eqref{timed} and \eqref{rhos} into \eqref{cyl-plane}, it is not difficult to find that the time derivative of the correlator can be expressed in terms of $\rho_i$ as
\be\label{dtG:rho}
{\frac{d}{dt}\log G_{E_L,E_R}(t)=\frac{c(\rho_2^2-\rho_1^2+\bar\rho_2^2-\bar\rho_1^2)}{6}}\ ,
\ee
with $E_L=-\frac{c\rho_1^2}{6},E_R=-\frac{c\bar\rho_1^2}{6}$.
The matrices $J_i$ are determined by junction conditions across the defects at $|z_i|=r_i$. Due to the presence of delta functions in $T(z)$, the solution $V(z)$ should be continuous across the defects but its first order derivative jumps. It is not difficult to show that the junction conditions are given by
\be
J_1V_1-V_2\big|_{|z|={r_1}}=J_3V_3-V_2\big|_{|z|={r_2}}=0\ ,
\ee
as well as
\be
V'_2-J_1V'_1-\frac{6{h_D}/c}{z}V_2\big|_{|z|={r_1}}=J_2V'_3-V'_2-\frac{6{h_D}/c}{z}V_2\big|_{|z|={r_2}}=0\  .
\ee
With these conditions, $J_i$ can be solved to be
\be\ba\label{solJ12}
&J_1=\frac{1}{2\rho_1}\left(\begin{array}{cc}
   z^{\rho_1-\rho_2}(\frac{6{h_D}}{c}+\rho_1+\rho_2)  &  z^{-\rho_1-\rho_2}(-\frac{6{h_D}}{c}+\rho_1-\rho_2) \\
  z^{\rho_1+\rho_2}(\frac{6{h_D}}{c}+\rho_1-\rho_2)    & z^{-\rho_1+\rho_2}(-\frac{6{h_D}}{c}+\rho_1+\rho_2) 
\end{array}\right)\ ,\quad |z|={r_1}\ ,
\\&J_2=\frac{1}{2\rho_1}\left(\begin{array}{cc}
   z^{\rho_1-\rho_2}(-\frac{6{h_D}}{c}+\rho_1+\rho_2)  &  z^{-\rho_1-\rho_2}(\frac{6{h_D}}{c}+\rho_1-\rho_2) \\
  z^{\rho_1+\rho_2}(-\frac{6{h_D}}{c}+\rho_1-\rho_2)    & z^{-\rho_1+\rho_2}(\frac{6{h_D}}{c}+\rho_1+\rho_2) 
\end{array}\right)\ ,\quad |z|={r_2}\ .
\ea
\ee
Again, there are anti-holomorphic parts which can be solved similarly. 
For the vacuum conformal block, the monodromy conditions suggest that the solutions $V$ and $\bar V$ should undergo trivial monodromies as they circle around a closed loop crossing the defects. 
Using  \eqref{sol-Fuchsian}, the monodromy matrix $\mathcal{M}$ is given by
\be\label{mat:mono}
\mathcal{M}=J_1(z_1)J_1^{-1}(z_2)J_2(z_3)J_2^{-1}(z_4)
\ee
with
\be
z_1=r_1e^{i\theta_1}\ ,\quad z_2=r_1e^{i\theta_2}\ ,\quad z_3=r_2e^{i\theta_2}\ ,\quad z_4=r_2e^{i\theta_1}\ .
\ee
Requiring $\mathcal{M}=\bar {\mathcal{M}}=\textbf{1}_{2\times 2}$, we get two monodromy equations 
\be\ba\label{mono:rho2}
&e^{2\rho_2(t_2-t_1)}=\frac{(\rho_2+\rho_1-6{h_D}/c)(\rho_2-\rho_1-6{h_D}/c)}{(\rho_2-\rho_1+6{h_D}/c)(\rho_2+\rho_1+6{h_D}/c)}\ ,
\\&e^{2\bar\rho_2(t_2-t_1)}=\frac{(\bar\rho_2+\bar\rho_1-6\bar{h}_D/c)(\bar\rho_2-\bar\rho_1-6\bar{h}_D/c)}{(\bar\rho_2-\bar\rho_1+6\bar{h}_D/c)(\bar\rho_2+\bar\rho_1+6\bar{h}_D/c)}\ .
\ea
\ee
Once solving these algebraic equations for $\rho_2$ and $\bar\rho_2$, the time derivative of the correlator can be obtained using \eqref{dtG:rho}. 

\subsection{Comparison between ETH and CFT$_2$}
Before computing the correlator from holographic side, we pause to 
compare the microcanonical correlators derived from the ETH ansatz and the monodromy method. It is easy to see from \eqref{Gplane} that the correlator can be factorized into holomorphic and anti-holomorphic parts with each being identical to one half of the corresponding scalar correlator.  More precisely, it can be shown that
\be\label{relation}
\log G_{E_L,E_R}(t)=\frac{1}{2}(\log G_{2E_L}(t)+\log G_{2E_R}(t))\ ,\ee
where $G_{2E_L}$ and $G_{2E_R}$ are scalar correlators defined in \eqref{correlator}.
Plugging \eqref{saddle:GEh} into \eqref{relation}, the right hand side can be written as \footnote{We have performed a change of the integrated variables $E'_a\to 2E'_a$ without altering the result. }
\be
\sum_{a=L,R}\frac{S(2E'_a)}{2}+t(E_a-E_a')-\frac{1}{2}(f_{h_D}(2E_L,2E_L')+f_{\bar h_D}(2E_R,2E_R'))\ ,
\ee
with $E_a'$ satisfying the saddle point equations. Using the fact that $S(2E)=2S(E/2)$ and comparing with \eqref{GELER}, we can easily get the envelop function for the spinning defect
\be
\tilde f(E_L,E_R,E_L',E_R')=\frac{f_{h_D}(2E_L,2E_L')+f_{\bar h_D}(2E_R,2E'_R)}{2}\ .
\ee
The factorization property of the microcanonical correlator \eqref{relation} also has important implications on the thermal correlator. Plugging \eqref{relation} into \eqref{def:G-spin}, we have
\be\ba\label{ther-fac}
Z_{\beta_L,\beta_R}G_{\beta_L,\beta_R}(t)&=\int dE_LdE_Re^{\tilde S(E_L,E_R)-\beta_LE_L-\beta_RE_R+\frac{1}{2}(\log G_{2E_L}(t)+\log G_{2E_R}(t))}
\\&=\int \frac{dE_LdE_R}{4}e^{\frac{1}{2}\left[S(E_L)+S(E_R)-\beta_LE_L-\beta_RE_R+\log G_{E_L}(t)+\log G_{E_R}(t)\right]}
\\&=[Z_{\beta_L}G_{\beta_L}(t)Z_{\beta_R}G_{\beta_R}(t)]^{\frac{1}{2}}\ ,
\ea
\ee
where in second line we changed variables $E_a\to E_a/2$, and in last line we used the definition for non-spinning thermal correlator \eqref{correlator}. The pure numerical factor can be ignored since we are interested in quantities of order $e^c$. As a result, the thermal correlator also factorizes.
Finally, using  \eqref{grav-act}, equation \eqref{ther-fac} can be rewritten as
\be 
\mathcal{I}=-\log[Z_{\beta_L,\beta_R}G_{\beta_L,\beta_R}(t)]=\mathcal{I}_L+\mathcal{I}_R\ , \label{factorize}
\ee
where 
\be\label{Ia}
\mathcal{I}_{a}=-\frac{1}{16G}(tM_{+a}+(\beta_{a}-t)M_{-a})-m_a\log\frac{r_{*a}}{2},\quad r_{*a}^2=M_{+a}+\left(\frac{M_{+a}-M_{-a}}{2m_a}-\frac{m_a}{2}\right)^2,
\ee
for $a=L,R$ and
\be
m_L=8Gh_D\ ,\quad m_R=8G\bar h_D\ .
\ee
 The masses $M_{\pm L,R}$ are subject to the conditions
\be\label{eqn}
\frac{r_{-a}}{r_{*a}}=\sin\frac{r_{-a}(\beta_a-t)}{2}\ ,\quad \frac{r_{+a}}{r_{*a}}=\sin\frac{r_{+a}t}{2},\quad a=L,R\ ,
\ee
with $r_{\pm a}=\sqrt{M_{\pm a}}$.

\section{Spinning Defect Correlators: AdS$_3$ Gravity}
In this section, we compute the correlation function of spinning defects in a rotating BTZ black hole background. 
In particular, we focus on the following thermal correlator
\be\label{thermal-G}
G_{\beta,\theta}(t)\equiv Z_{\beta,\theta}\langle D^\dagger(t)D(0)\rangle_{\beta,\theta}\ ,
\ee
where $\beta$ is the inverse temperature, and $\theta$ is the angular potential. They are related to the left and right moving temperatures introduced in \eqref{def:G-spin} via
\be\label{beta-theta}
\beta_L=\beta+i\theta\ ,\quad\beta_R=\beta-i\theta\ .
\ee
We compute \eqref{thermal-G} by evaluating the Euclidean on-shell action of the backreacted geometry,
which is composed of two rotating BTZ black holes glued across the domain wall via proper junction condition. If the defect operator is spinless, we impose Israel's junction condition \eqref{jun1}. The situation changes when we consider spinning defects and the junction condition becomes a little subtle. A careful analysis using the first order formalism   suggests that the induced metric should also be discontinuous across the   domain wall. In the following, we will study the backreacted geometries and
compute the thermal correlators for both non-spinning and spinning defect operators.

\subsection{Non-spinning defects}
We first consider rotating BTZ black hole backreacted by a spherical shell of non-spinning dust particles. This is similar to  the case of gluing a rotating BTZ black hole with the vacuum AdS studied in \cite{Chandra:2024vhm}. In our case,
the dust divides the spacetime into two rotating BTZ black holes with metrics in each region being
\be\label{rotatingbtz}
ds^2_\pm=f_\pm(r)dt_\pm^2+\frac{dr_\pm^2}{f_\pm (r)}+r_\pm^2\left(d\phi_\pm-i\frac{J_\pm}{2r^2}dt_\pm\right)^2\ ,
\ee
where the blackening factor $f_\pm$ is given by
\be
f_\pm=r^2_\pm-M_\pm+\frac{J_\pm^2}{4r_\pm^2}\ .
\ee 
The metrics $ds^2_-$ and $ds^2_+$ describe the geometries to the left and right of the domain wall respectively. Since we are considering the Euclidean BTZ black holes, we let $J_\pm$ be pure imaginary.  Instead of studying the dust as a whole directly, we first analyze the motion of a single particle   for later convenience. Since discussions in the two regions are parallel, we omit the subscript $\pm$ in the following for  simplicity. Let the geodesic be parametrized by $(t(\ell),r(\ell),\phi(\ell))$ and $v=(t'(\ell),r'(\ell),\phi'(\ell))$ be the velocity of the particle with  $\ell$ being the affine parameter.  The geodesic equation can be recast into three first order ODEs which are given by
\be
(\p_{t})_\mu v^\mu=e\ ,\quad (\p_{\phi})_\mu v^\mu=j\ ,\quad v^2=1\ ,
\ee
where $e$ and $j$ are constants parametrizing the energy and angular momentum of the particle. Solving these equations gives
\be\ba\label{geo}
&t'(\ell)=\frac{4er^2+2iJj}{4(r^2-r_o^2)(r^2-r_i^2)}\ ,
\\&r'(\ell)=\pm\frac{\sqrt{4r^4-4(M+j^2+e^2)r^2+J^2+4j^2M-4iejJ}}{2r}\ ,\\
&\phi'(\ell)=\frac{4jr^2-4jM+2ieJ}{4(r^2-r_o^2)(r^2-r_i^2)}\ ,
\ea\ee
where $r_o$ and $r_i$ are outer and inner horizon radii given by
\be
r_o=\sqrt{\frac{M+\sqrt{M^2-J^2}}{2}}\ ,\quad r_i=\sqrt{\frac{M-\sqrt{M^2-J^2}}{2}}\ .
\ee
Due to rotational symmetry, energy $e$ and angular momentum $j$ should be same for all particles in each region. Collecting all particles and letting $\ell,\psi$ be the coordinates on the domain wall such that $dt=t'd\ell,dr=r'd\ell,d\phi=d\psi+\phi'd\ell$, then the induced metric on the two sides of the domain wall can be computed using \eqref{geo} and the result is given by
\be\label{ind}
ds^2\Big|_{\mathcal W}=\left(1-\frac{j^2}{r(\ell)^2}\right)d\ell^2+r(\ell)^2\left(d\psi+\frac{j}{r(\ell)^2}d\ell\right)^2\ .
\ee
By Israel's junction condition \eqref{jun1}, 
the continuity of the induced metric implies that $j_+=j_-$ and $r(\ell)$ is continuous across the shell. Using \eqref{geo}, the continuity of $r'(\ell)$ implies
\be\label{eq1}
[M+e^2]=[J-2iej]=0\ .
\ee
The normal vector to the worldvolume of the dust should be orthogonal to both $\p_\ell$ and $\p_\psi$, and is therefore given by
\be\label{normal}
n= \left(1-\frac{j^2}{r^2}\right)^{-\frac{1}{2}}(-r'dt+t'dr)\ ,\quad n^2=1\ .
\ee
Using \eqref{geo}, we can compute the extrinsic curvature and obtain
\be\ba
& K_{\ell\ell}=0\ ,\quad K_{\ell\psi}=\frac{2ej+iJ}{2\sqrt{r^2-j^2}}\ ,\quad K_{\psi\psi}=\frac{iJj+2er^2}{2\sqrt{r^2-j^2}}\ .
\ea\ee
The stress tensor can be determined from the second equation of the junction condition \eqref{jun1} with the result being
\be\label{stress-2d}
T^{\ell\ell}=[-\frac{2er^2+ijJ}{2(r^2-j^2)^{3/2}}]=-\frac{[e]}{\sqrt{r^2-j^2}}\ ,\quad T^{\ell\psi}= {T^{\psi\psi}}=0\ .
\ee
It is easy to check that the stress tensor is conserved $\nabla_aT^{ab}=0$. We define $m$ as the total mass of the dust observed by an asymptotic observer so that $T^{\ell\ell}\sim 4Gmr^{-1}$. Using  \eqref{stress-2d}, this implies
\be\label{eq2}
[e]=-4Gm\ .
\ee
Combining \eqref{eq1} with \eqref{eq2}, we find 
\be\ba\label{nospin}
e_\pm=\frac{M_--M_+\pm m ^2}{2m}\ ,\quad j_\pm=\frac{i(J_--J_+)}{2m}\ .
\ea\ee
Therefore, the geodesic is completely determined in terms of the data of the two black holes and total mass of the dust.
Plugging \eqref{nospin} into \eqref{geo}, we get
\be
(r^2)'(\ell)=\pm2\sqrt{(r^2-r^2_{*1})(r^2-r^2_{*2})}\ ,
\ee
where the turning points $r_{*i}$ are given by
\be
r_{*i}^2=\frac{M+j^2+e^2+(-1)^i\sqrt{\mathcal{A}}}{2}\ ,\quad i=1,2
\ee
with
\be\ba
\mathcal A=&(2m)^{-4}(m^2+(r_{i+}+r_{i-}-r_{o+}-r_{o-})^2)(m^2+(r_{i+}-r_{i-}-r_{o+}+r_{o-})^2)
\\&\times(m^2+(r_{i+}-r_{i-}+r_{o+}-r_{o-})^2)(m^2+(r_{i+}+r_{i-}+r_{o+}+r_{o-})^2)\ .
\ea\ee
It can be checked that $r_{*2}>r_{o\pm}$ and $r_{*1}^2<0$. Furthermore, we can compute the time elapsed by the domain wall which will be useful later
\be
\ba\label{delta-t}
\Delta t_\pm=2\int_{r_{*2}}^\infty dr\frac{dt}{dr}=\frac{\beta_\pm}{\pi}\arctan\sqrt{\frac{r^2_{*1}-r^2_{o\pm}}{r^2_{o\pm}-r^2_{*2}}}+\frac{i\theta_\pm}{\pi}\arctan\sqrt{\frac{r^2_{*1}-r^2_{i\pm}}{r^2_{i\pm}-r^2_{*2}}}\ ,
\ea
\ee
where
\be\label{beta-r}
\beta_\pm=\frac{2\pi r_{o\pm}}{r^2_{o\pm}-r^2_{i\pm}}\ ,\quad \theta_\pm=\frac{2\pi i r_{i\pm}}{r^2_{o\pm}-r^2_{i\pm}}\ .
\ee
Similarly, the angular shifted by the shell is given by
\be
\ba\label{delta-phi}
\Delta \phi_\pm=\frac{\theta_\pm}{\pi}\arctan\sqrt{\frac{r^2_{*1}-r^2_{o\pm}}{r^2_{o\pm}-r^2_{*2}}}-\frac{i\beta_\pm}{\pi}\arctan\sqrt{\frac{r^2_{*1}-r^2_{i\pm}}{r^2_{i\pm}-r^2_{*2}}}\ .
\ea
\ee
\subsubsection{On-shell action}
Having determined the trajectory of the domain wall, we are about to compute the on-shell action of the saddle point geometry. We assume that the boundary insertion time $t$ satisfies $t<\beta/2$ without loss of generality.  This guarantees that the domain wall is always on the right  side of the Euclidean disk of the left black hole. As we increase the value of $t$ from $0$ to $\beta/2$, there are three possible configurations of the trajectory of dust in the right black hole. Let us consider a domain wall which starts from the boundary at time $-t/2$, moves into the bulk and returns back to the boundary at time $t/2$ with $t\leq t_{c_1}$ for some critical value $t_{c_1}$. This trajectory would feature only one turning point at the $t=0$ slice, and the trajectory remains to the right part of the disk.
As the insertion time $t$ increases, the trajectory gets closer to the horizon of the right black hole. Due to the influence of the black hole, the trajectory would have additional turning points. But it still sits to the right hand side.
As $t$ gets further increased such that $t\geq t_{c_2}$ for the second critical time $t_{c_2}$, the trajectory moves into the opposite side of the Euclidean disk, see figure 14 in \cite{Chandra:2024vhm} for more details.

\begin{figure} 
    \centering
    \includegraphics[width=0.44\linewidth]{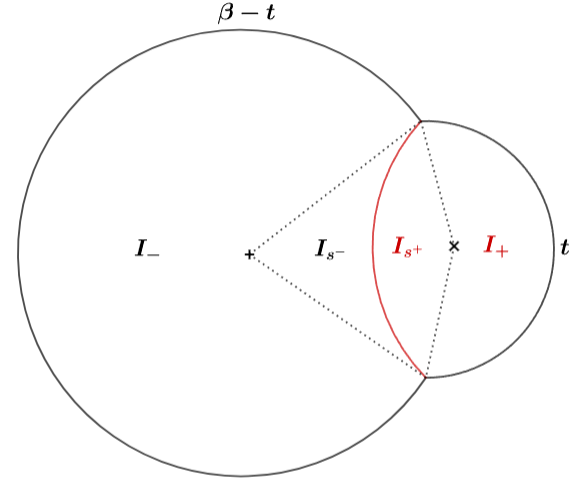}
    \hfill
     \includegraphics[width=0.4\linewidth]{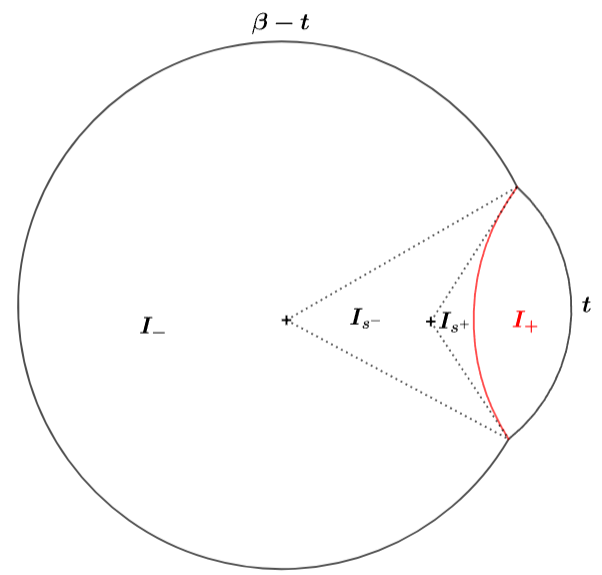}
    \caption{Backreacted geometry for $t>t_{c_2}$(left) and $t\leq t_{c_1}$(right). Red lines denote worldvolumes of dust or domain walls $\mathcal{W}$. The whole spacetime is divided into four regions with the action in each region denoted by $I_-,I_+, I_{s^-}$ and $I_{s^+}$.}
    \label{figsect3}
\end{figure}

The gravitational action is again given by \eqref{grapre}, which we repeat below for convenience
\be\label{action}
I=-\frac{1}{16\pi G}\int\sqrt{g}(R+2)+\int_{\mathcal{W}}\sqrt{h}\sigma\ .
\ee
Taking the trace of the  Einstein equations, we find
\be\label{eins}
R+6=16\pi G\sigma {\delta(\mathcal{W})}\ .
\ee
Substituting \eqref{eins} into \eqref{action}, we find that the contribution from the shell action cancels against the contribution from delta function in the Einstein-Hilbert action. As a result, the on-shell action we need to evaluate is given by
\be
I=\frac{1}{4\pi G}\int\sqrt{g}\ .
\ee
In the following, we will compute the Euclidean on-shell action of the backreacted geometry. In particular, we will discuss the case when $t>t_{c_2}$ and $t\leq t_{c_1}$ separately. The intermediate case $t_{c_1}<t<t_{c_2}$ is similar to the case $t\leq t_{c_1}$  since the shell remains to the right side of the Euclidean disk. Evaluating the action for $t_{c_1}<t<t_{c_2}$ is more involved but the result is expected to be the same. Actually, the saddle point equations and on-shell actions in different cases are related by proper analytic continuations.
\paragraph{Case 1: $t>t_{c_2}$} 
We begin with the simpler case with $t_{c_2}<t$. To evaluate the action, we can divide the spacetime into four regions as shown in the left of figure \ref{figsect3}.
Correspondingly, the action can be expressed as a summation of four terms
\be\label{separate}
I=I_-+I_++I_{s^-}+I_{s^+}\ .
\ee
The actions $I_\pm$ can be obtained easily since they are part of the BTZ action and we find
\be\label{Ipm}
I_+=-\frac{t(r_{o+}^2-r_{i+}^2)}{ {8G}}\ ,\quad I_-=-\frac{(\beta-t)(r_{o-}^2-r_{i-}^2)}{ {8G}}\ .
\ee
The action $I_{s^-}$ is given by 
\be\ba\label{Ism}
I_{s^-}&=\frac{1}{G}\int_0^{t/2} d\bar t\int_{r_{o-}}^{r(\bar t)} drr=\frac{1}{2G}\int^{r_\infty}_{r_{*2}}dr\frac{dt_-}{dr}(r^2-r^2_{o-})
\\&=-\frac{e_-}{2G}\log\frac{2r_\infty}{\sqrt{r^2_{*2}-r^2_{*1}}}+\frac{r_{i-}}{2G}\arctan\sqrt{\frac{r^2_{*1}-r^2_{i-}}{r^2_{i-}-r^2_{*2}}}\ ,
\ea\ee
where $r_\infty$ is the boundary cutoff.
Similarly $I_{s^+}$ is\footnote{The minus sign in the second equality arises because $r(t/2)=r_\infty,r(\beta/2)=r_{*2}$.} 
\be\ba\label{Isp}
I_{s^+}&=\frac{1}{G}\int_{t/2}^{\beta/2} d\bar t \int_{r_{o+}}^{r(\bar t)} drr=-\frac{1}{2G}\int^{r_\infty}_{r_{*2}}dr\frac{dt_+}{dr}(r^2-r^2_{o+})
\\&=\frac{e_+}{2G}\log\frac{2r_\infty}{\sqrt{r^2_{*2}-r^2_{*1}}}+\frac{r_{i+}}{2G}\arctan\sqrt{\frac{r^2_{*1}-r^2_{i+}}{r^2_{i+}-r^2_{*2}}}\ .
\ea\ee
Therefore, the action of the intermediate region is given by the sum of above actions
\be
\ba\label{Is}
I_s&\equiv I_{s^-}+I_{s^+}
\\&=\frac{m}{2G}\log\frac{2r_\infty}{\sqrt{r^2_{*2}-r_{*1}^2}}+\frac{r_{i-}\arctan\sqrt{\frac{r^2_{*-}-r^2_{i-}}{r^2_{i-}-r^2_{*+}}}+r_{i+}\arctan\sqrt{\frac{r^2_{*-}-r^2_{i+}}{r^2_{i+}-r^2_{*+}}}}{2G}
\\&=\frac{m}{2G}\log\frac{2r_\infty}{\sqrt{r^2_{*2}-r_{*1}^2}}+\frac{ir_{i-}r_{o-}\Delta\phi_-+ir_{i+}r_{o+}\Delta\phi_++r_{i-}^2\Delta t_-+r_{i+}^2\Delta t_+}{4G}\ ,
\ea
\ee
where $\Delta t_\pm$ and $\Delta\phi_\pm$ are given by \eqref{delta-t} and \eqref{delta-phi}.
We choose the   counterterm to be
\be\label{Ict}
I_{\text{ct}}=-\frac{m}{2G}\log r_\infty
\ee
to remove the logarithmic divergence. As a result, the thermal correlation function for non-spinning thin-shell operators is given by
\be\label{cor}
G_{\beta,\theta}(t)\approx e^{-\Delta I}\ ,\quad \Delta I=I+I_{\text{ct}}+\log Z_{\beta,\theta}
\ee
at the leading order in $G^{-1}$ expansion.
Moreover, the time elapsed $\Delta t_\pm$ by the domain wall should satisfy
\be\label{eqt1}
\beta_-=\beta-t+\Delta t_-\ ,\quad \beta_+=\Delta t_++t\ .
\ee
Similarly, $\Delta\phi_\pm$ should satisfy
\be\label{eqp1}
\theta_-=\theta+\Delta\phi_-\ ,\quad\theta_+=\Delta\phi_+\ .
\ee
Plugging \eqref{eqt1}  and \eqref{eqp1} into \eqref{Is}, we get
\be
I_s+I_{\text{ct}}=\frac{m}{2G}\log\frac{2}{\sqrt{r^2_{*2}-r^2_{*1}}}-\frac{tr^2_{i+}+(\beta-t)r^2_{i-}+ir_{i-}r_{o-}\theta}{4G}\ .
\ee
Combining with \eqref{Ipm},  the total action is given by
\be\label{final}
I+I_{\text{ct}}=\frac{m}{2G}\log\frac{2}{\sqrt{r^2_{*2}-r^2_{*1}}}-\frac{t(r^2_{o+}+r^2_{i+})+(\beta-t)(r^2_{o-}+r^2_{i-})+2ir_{i-}r_{o-}\theta}{8G}\ .
\ee
To compute the correlation function, we need to solve \eqref{eqt1} and \eqref{eqp1} for $(M_\pm, J_\pm)$ and substitute the solution into \eqref{final}.

\paragraph{Case 2: $t\leq t_{c_1}$} In this case, the domain wall locates on the right part of the Euclidean disk of both black holes as can be seen in figure \ref{figsect3}. 
The total action can be expressed as
\be
I=I_-+I_++I_{s^-}-\tilde I_{s^+}\ ,
\ee
where $I_\pm$ and $I_{s^+}$ have the same expression as before. $\tilde I_{s^+}$ is given by
\be
\tilde I_{s^+}=\frac{1}{G}\int_0^{t/2} d\bar {t/2}\int_{r_{o+}}^{r(\bar t)} drr=\frac{1}{2G}\int^{r_\infty}_{r_{*2}}dr\frac{dt_+}{dr}(r^2-r^2_{o+})
=-I_{s^+}\ .
\ee
Therefore, the total action in this case has the same expression as the first case when $t>t_{c_2}$.   Actually, the second case can be considered as an analytically continuation of the first case. The saddle point equations satisfied by $M_\pm, J_\pm$ are now
\be
\beta_-=\beta-t+\Delta t_-\ ,\quad \Delta t_+=t\ ,
\ee
and
\be\label{eqp2}
\theta_-=\theta+\Delta\phi_-\ ,\quad \Delta\phi_+=0\ .
\ee

\subsubsection{Matching}
Having computed the on-shell action \eqref{final}, we can now prove the thermal correlator agrees with that computed from field theory side. The proof consists of two steps. First, we show that with $m_L=m_R=m$, the left and right moving energies $M_{\pm L},M_{\pm R}$ introduced in \eqref{Ia} in field theory are identical to $ {M_\pm+J_\pm} $ and $ {M_\pm-J_\pm}$ respectively by showing they satisfy the same equations. Then we will show that $\mathcal{I}$ given by \eqref{factorize} is exactly the gravitational on-shell action.

 We define
\be\ba\label{Mlr}
 M_{\pm l}=  { r_{\pm l}^2}\equiv{M_\pm+J_\pm}\ ,\quad  M_{\pm r}= { r_{\pm r}^2}\equiv {M_\pm-J_\pm}\ .
\ea\ee
To show that $ M_{\pm l,r}$  satisfy the same equations as $M_{\pm L,R}$  which are
given by \eqref{eqn}, we rewrite gravitational saddle point equations \eqref{eqt1}  and \eqref{eqp1} using their linear combinations
\be
\ba\label{eqns}
&(\beta_-+ i\theta_-)\left(1-\frac{\arcsin\frac{ r_{-l}}{ r_{*l}}}{\pi}\right)=\beta+ i\theta-t\ ,\quad (\beta_++ i\theta_+)\left(1-\frac{\arcsin\frac{ r_{+l}}{ r_{*l}}}{\pi}\right)=t\ ,
\\&(\beta_-- i\theta_-)\left(1-\frac{\arcsin\frac{ r_{-r}}{ r_{*r}}}{\pi}\right)=\beta- i\theta-t\ ,\quad (\beta_+- i\theta_+)\left(1-\frac{\arcsin\frac{ r_{+r}}{r_{*r}}}{\pi}\right)=t\ ,
\ea
\ee
where  $ r_{*l,r}$ has the same expression as $r_{*L,R}$ in \eqref{Ia}, but with $M_{\pm L,R}$ replaced by $ M_{\pm l,r}$ and $m_L=m_R=m$. 
To get \eqref{eqns}, we have used the following identities
\be\ba\label{identity}
&\arcsin\frac{ r_{\pm l}}{ r_{*l}}=\arctan\sqrt{\frac{r_{*1}^2-r_{o\pm}^2}{r_{o\pm}^2-r_{*2}^2}} +\arctan\sqrt{\frac{r_{*1}^2-r_{i\pm}^2}{r_{i\pm}^2-r_{*2}^2}} \ ,
\\&\arcsin\frac{r_{\pm r}}{r^*_r}=\arctan\sqrt{\frac{r_{*1}^2-r_{o\pm}^2}{r_{o\pm}^2-r_{*2}^2}} - \arctan\sqrt{\frac{r_{*1}^2-r_{i\pm}^2}{r_{i\pm}^2-r_{*2}^2}} \ .
\ea\ee
Combining \eqref{beta-r} with \eqref{Mlr}, we have
\be
\beta_\pm+ i\theta_\pm=\frac{2\pi}{r_{o\pm}+r_{i\pm}}=\frac{2\pi}{ r_{\pm l}}\ ,\quad \beta_\pm- i\theta_\pm=\frac{2\pi}{r_{o\pm}-r_{i\pm}}=\frac{2\pi}{r_{\pm r}}\ .
\ee
Substituting this into \eqref{eqns} gives
\be
\ba
&\pi-\arcsin\frac{ r_{-l}}{ r_{*l}}=\frac{ r_{-l}(\beta+i\theta-t)}{2}\ ,\quad  \pi-\arcsin\frac{ r_{+l}}{  r_{*l}}=\frac{r_{+l}t}{2}\ ,
\\&\pi-\arcsin\frac{ r_{-r}}{ r_{*r}}=\frac{ r_{-r}(\beta-i\theta-t)}{2}\ ,\quad \pi-\arcsin\frac{ r_{+r}}{ r_{*r}}=\frac{ r_{+r}t}{2}\ .
\ea
\ee
These equations match exactly with \eqref{eqn} using the relation \eqref{beta-theta}. Since $M_{\pm l,r}$ and $M_{\pm L,R}$ satisfy same equations, we can simply identify them and get
\be \label{match}
r_{*l}=r_{*L}\ ,\quad r_{*r}=r_{*R}\ .
\ee With such identifications, 
it can be further checked directly that
\be
r_{*L}r_{*R}=r_{*2}^2-r_{*1}^2\ .
\ee 
Consequently, the expected factorized form of gravitational action \eqref{factorize} from the ETH analysis and monodromy method can be expressed in terms of variables in gravity as
\be\label{Is0}
\log [Z(\beta)G_{\beta_L,\beta_R}(t)]\approx \frac{m}{2G}\log\frac{2}{\sqrt{r^2_{*2}-r^2_{*1}}}-\frac{t(r^2_{o+}+r^2_{i+})+(\beta-t)(r^2_{o-}+r^2_{i-})+2ir_{i-}r_{o-}\theta}{8G}\ .
\ee
This is exactly the same result as \eqref{final}.  Thus we finish the proof of factorization of on-shell gravitational action, and the matching between three parts. Although we have focused on the first case where $t>t_{c_2}$, the factorization of the second case at early time is straightforward to prove in the same way since it can be viewed as a analytical continuation of the first case.

\subsection{Spinning Defects}

\subsubsection{Junction condition}
Having computed the thermal correlator of non-spinning defects in  the rotating BTZ black hole, we are now ready to move to the more general case for spinning defects.
We  start with studying the motion of the trajectory of a single rotating dust as before. Generally speaking, the velocity $v^\mu$ of a spinning particle with mass $m$ and spin $s$ in AdS$_3$ obeys the MPD equation \cite{Mathisson:1937zz,Papapetrou:1951pa,Dixon:1970zza}
\be\label{MPD}
\nabla(mv^\mu-s\epsilon^{\mu\nu\lambda}v_\nu\nabla v_\lambda)=0\ ,
\ee
which is a third order differential equation.  It is clear that the geodesic  $\nabla v^\mu=0$ is a solution to the MPD equation. On the other hand,  notice that if $v^\mu$ allows a perturbative expansion in small $s$
\be
v=v^{(0)}+sv^{(1)}+O(s^2)\ ,
\ee
equation \eqref{MPD} would imply that $\nabla v=0$ holds to all orders as long as $\nabla v^{(i)}$ is finite in $s$. Therefore, additional solutions of the MPD equation would have non-trivial "non-perturbative" effects in $s$ and we will not investigate them here. Consequently, the trajectory is still described by \eqref{geo} and the induced metric on domain wall is also given by \eqref{ind}. However, we need to relax the junction condition by allowing the induced metric to be discontinuous across the domain wall. As a result, $j_+\neq j_-$. Before giving the precise junction condition, we provide one evidence that
supports the discontinuity feature. We note that the stress tensor of a spinning particle is given by \cite{Castro:2014tta}
\be\label{Tspin}
T^{\mu\nu}_{\text{spin}}=\int\textbf{dl}[mv^\mu v^\nu+iv_\alpha v^{(\mu}\nabla s^{\nu)\alpha}]+i\nabla_\alpha\left(\int \textbf{dl}v^{(\mu}s^{\nu)\alpha}\right)\ ,
\ee
where $s^{\mu\nu}$ is the spin tensor. Let $n_1$ and $n_2$ be two unit normal vectors satisfying
\be
n_1^2=n_2^2=1\ ,\quad n_1\cdot v=n_2\cdot v=n_1\cdot n_2=0\ ,
\ee
then $s^{\mu\nu}$ is given by
\be
s^{\mu\nu}=sn_1^{[\mu}n_2^{\nu]}\ .
\ee
From \eqref{Tspin}, we can see that the stress tensor contains a term proportional to the derivative of the delta function. After summing the contribution from  all particles by integrating over $\phi$, the stress tensor of the domain wall can be formally written as
\be\label{T-spin}
T^{ab}=\mathcal{T}^{ab}\delta(\mathcal{W})+\mathcal{S}^{ab}\delta'(\mathcal{W})\ .
\ee
On the other hand, we can also formally write the metric of the whole spacetime as
\be
ds^2=ds_+^2\Theta^+(\mathcal W)+ds_-^2\Theta^-(\mathcal W)\ .
\ee
where $\Theta^+(\Theta^-)$ takes value of  1 to the right (left) hand side of the worldvolume $\mathcal{W}$ and 0 in the opposite region. If the induced metric is continuous so that $[ds^2]=0$, then it can be shown that Einstein equations can only reproduce the first term in \eqref{T-spin}, which amounts to the second junction condition in \eqref{jun1}. As a result, one has to give up the requirement of continuity in order to satisfy the Einstein equations on rotating domain wall. 

Actually, the spinning object will source a nontrivial torsion and it is more convenient to study its backreaction using the first order formalism where the vielbeins and spin connection are treated independently \cite{Watanabe:2004nt}.
The backreaction of rotating domain wall in three dimensional spacetime has already been studied in \cite{Giacomini:2006um}
and it was indeed found that instead of  a continuous metric, we should require
\be\label{require}
[\text{Vol}]=0\ ,\quad \iota[\mathrm{e}^{a}\wedge [\mathrm{e}^{b}]]=0\ .
\ee
The first equation means that the volume form of the induced metric is continuous across the domain wall.  The second equation is a constraint on the vielbeins $\mathrm{e}^a$ on both sides of the domain wall with $\iota$ denoting the pullback of differential forms to the domain wall. Conditions \eqref{require} are imposed in order to make the stress tensor $T^\mu_a$ and spin current $S^\mu_{ab}$ well defined even without a continuous induced metric.
By virtue of \eqref{ind}, this suggests that the radial coordinates on two sides of domain wall are different. The continuity of the volume form suggests the following relations on domain wall
\be\label{rp-rm}
r_+^2-j_+^2=r_-^2-j_-^2\equiv \tilde r^2\ .
\ee
The second condition of \eqref{require}  is also satisfied which can be easily seen by properly choosing the veilbeins. We show the details in appendix \ref{Appendix}.  Since \eqref{rp-rm} implies that $(\tilde r^2)'$ is   continuous, equations \eqref{eq1} still hold. 
Defining $s$ as
\be
s\equiv -i(j_+-j_-)\ ,
\ee
and solving \eqref{eq1} and \eqref{eq2}, we get
\be\label{ej:spin}
e_\pm=\frac{m(M_--M_+)-s(J_--J_+)}{2(m^2-s^2)}\pm\frac{m}{2}\ ,\quad j_\pm=\frac{i[m(J_--J_+)-s(M_--M_+)}{2(m^2-s^2)}\pm\frac{is}{2}\ .
\ee
Having determined the parameters of geodesic, we need to calculate the time elapse $\Delta t_{\pm}$ and angular shift $\Delta \phi_{\pm}$. Plugging \eqref{ej:spin} into \eqref{geo}, we find the radial coordinate of the geodesic satisfies
\be
(\tilde r^2)'=\pm 2\sqrt{(\tilde r^2-\tilde r_{*1})(\tilde r^2-\tilde r^2_{*2})}\ ,
\ee
where the turning points $\tilde r_{*i}$ are  now modified to be
\be
\ba\label{tilders}
\tilde r^2_{*i}=\frac{M-j^2+e^2+(-1)^i\sqrt{\tilde{\mathcal{A}}}}{2}\ ,\quad i=1,2
\ea
\ee
with
\be\ba
\tilde{\mathcal A}=&(4(m^2-s^2))^{-2}((m-s)^2+(r_{i+}+r_{i-}-r_{o+}-r_{o-})^2)((m-s)^2+(r_{i+}-r_{i-}-r_{o+}+r_{o-})^2)
\\&\times((m+s)^2+(r_{i+}-r_{i-}+r_{o+}-r_{o-})^2)((m+s)^2+(r_{i+}+r_{i-}+r_{o+}+r_{o-})^2)\ . \nn
\ea\ee
Then the time elapsed and angular shifted by the rotating domain wall are 
\be\ba
&\Delta t_\pm = \frac{\beta_\pm}{\pi}\arctan\sqrt{\frac{\tilde r^2_{*1}-\tilde r^2_{o\pm}}{\tilde r^2_{o\pm}-\tilde r^2_{*2}}} + \frac{i\theta_\pm}{\pi}\arctan\sqrt{\frac{\tilde r^2_{*1}-\tilde r^2_{i\pm}}{\tilde r^2_{i\pm}-\tilde r^2_{*2}}} \
,
\\&\Delta \phi_\pm = \frac{\theta_\pm}{\pi}\arctan\sqrt{\frac{\tilde r^2_{*1}-\tilde r^2_{o\pm}}{\tilde r^2_{o\pm}-\tilde r^2_{*2}}} - \frac{i\beta_\pm}{\pi}\arctan\sqrt{\frac{\tilde r^2_{*1}-\tilde r^2_{i\pm}}{\tilde r^2_{i\pm}-\tilde r^2_{*2}}}
\ ,
\ea
\ee
with
\be
\tilde r_{o\pm}^2=r^2_{o\pm}-j_\pm^2\ ,\quad \tilde r_{i\pm}^2=r^2_{i\pm}-j_\pm^2\ ,
\ee
and $\beta_\pm$, $\theta_\pm$ are defined in \eqref{beta-r}.

\subsubsection{On-shell action}
Now we are ready to compute the on-shell gravitational action backreacted by rotating domain wall, which contains an additional spin term,
\be
I_{\text{tot}}=I+I_{\text{ct}}+I_{\text{spin}}\ ,
\ee
where 
\be\label{Ispin}
I_{\text{spin}}=\frac{s}{4G}\int d\psi d\ell \ , n_2^\mu v^\nu \nabla_\nu (n_1)_{\mu}\  .
\ee
For simplicity, we will focus on the case $t>t_{c_2}$. The action $I$ can be separated in the same way as \eqref{separate}, where actions $I_+$ and $I_-$ are still given by \eqref{Ipm}. We can follow the same routine to compute the middle part $I_s=I_{s^-}+I_{s^+}$: evaluating the integral explicitly and rewriting the results in terms of $\Delta t_\pm$ and $\Delta\phi_\pm$. Finally, using the saddle equations \eqref{eqt1}, \eqref{eqp1} and combining with the same counterterm as \eqref{Ict}, it can be checked that $ I+I_{\text{ct}}$ is the same as \eqref{final} with $r_{*i}$ replaced by $\tilde r_{*i}$ defined in \eqref{tilders}. We are therefore left with the last spinning action $I_{\text{spin}}$.  
We choose the normal vectors satisfying \eqref{normal} to be
\be
(n_1)_\mu dx^\mu=\frac{r}{\tilde r}(v^rdt-v^tdr)\ ,\quad (n_2)_\mu dx^\mu=\frac{2ej+iJ}{2\tilde r}dt+\frac{4jv^r}{f(r)\tilde r}dr-\tilde rd\phi\ ,
\ee
where $v^t=t'(\ell),v^r=r'(\ell)$.
Then the integrand of \eqref{Ispin} can be computed to be
\be
 n_2^\mu v^\nu \nabla_\nu (n_1)_{\mu}=\frac{i[J-2iej]}{2\tilde r^2}\ ,
\ee
which is single valued on the two sides of the shell  due to \eqref{eq1}. Computing the integral directly gives
\be
I_{\text{spin}}=-\frac{s}{2G}\tanh^{-1}\frac{\tilde r_{*2}}{\tilde r_{*1}}=-\frac{s}{4G}\log\frac{\tilde r_{*2}+\tilde r_{*1}}{\tilde r_{*2}-\tilde r_{*1}}\ .
\ee

\subsubsection{Matching}

Matching the gravitational on-shell action with the thermal correlator derived from  field theory for spinning defects can be done in a similar way as before. First,
we can check that the following identities similar to \eqref{identity} still hold
\be\ba\label{identity2}
&\arcsin\frac{ r_{\pm l}}{ r_{*l}}=\arctan\sqrt{\frac{\tilde r_{*-}^2-\tilde 
 r_{o\pm}^2}{\tilde r_{o\pm}^2-\tilde r_{*+}^2}} +\arctan\sqrt{\frac{\tilde r_{*-}^2-\tilde r_{i\pm}^2}{\tilde r_{i\pm}^2-\tilde r_{*+}^2}} \ ,
\\&\arcsin\frac{r_{\pm r}}{r^*_r}=\arctan\sqrt{\frac{\tilde r_{*-}^2-\tilde r_{o\pm}^2}{\tilde r_{o\pm}^2-\tilde r_{*+}^2}} - \arctan\sqrt{\frac{\tilde r_{*-}^2-\tilde r_{i\pm}^2}{\tilde r_{i\pm}^2-\tilde r_{*+}^2}}\ ,
\ea\ee
where $r_{*l,r}$ is given by \eqref{Ia} with $M_{\pm L,R}$ replaced by $M_{\pm l,r}$, and $m_L=m+s,m_R=m-s$. Using these identities, it is not difficult to show that $r_{\pm l,r}$ satisfies the same equation as $r_{\pm L,R}$. As a result, \eqref{match} still holds. Then the thermal correlation function computed from the ETH ansatz and field theory \eqref{factorize} can be rewritten as
\be\label{predict-G-s}
\mathcal{I}=I+I_{\text{ct}}-\frac{s}{4G}\log\frac{r_{*L}}{r_{*R}}\ .
\ee
In the end, it is not difficult to show 
\be
\frac{\tilde r_{*2}+\tilde r_{*1}}{\tilde r_{*2}-\tilde r_{*1}}=\frac{r_{*L}}{r_{*R}}\ .
\ee
Therefore, $I_{\text{spin}}=-\frac{s}{4G}\log\frac{r_{*L}}{r_{*R}}$, which is the same as the last term in \eqref{predict-G-s}. As a result, we show the gravitational action indeed factorize as expected and have
\be
\log[Z_{\beta_L,\beta_R}G_{\beta_L,\beta_R}(t)]=-I_{\text{tot}}=-\mathcal{I}\ .
\ee

\section{Four and Higher Point Correlators}

In this section, we extend results of \cite{Chen:2024hqu} to the cases including multiple line defects. First, we explicitly present the generalizations of the three corners, which are ETH, field theory and gravity, and  demonstrate how they match in the case of four point correlators. Then by summarizing features of four defect example, we provide a general picture for computing higher defect correlators. Due to the factorization property of the correlators, we work with the scalar line defects in this section for simplicity. Note the setup remains within the single boundary AdS/CFT correspondence instead of  multiboundary version with higher topologies. A general feature of multidefect insertions is that they should pair with each other to have non-vanishing solutions at the leading order in the large $N$ expansion. Besides, we get different results for different orders of defect insertions. {The leading order results of higher point defect correlators are consistent with Gaussian statistics as we will see soon. This suggests that in order to see non-Gaussian statistics, we have to go beyond the leading order result. On the field theory side, this means to study contributions from $1/c$ corrections to the vacuum block and other non-vacuum blocks. On the gravity side, non-Gaussian statistics can be explored by studying 
wormhole geometry with multi-boundaries}

\begin{figure}
    \centering
    \includegraphics[width=0.55\textwidth]{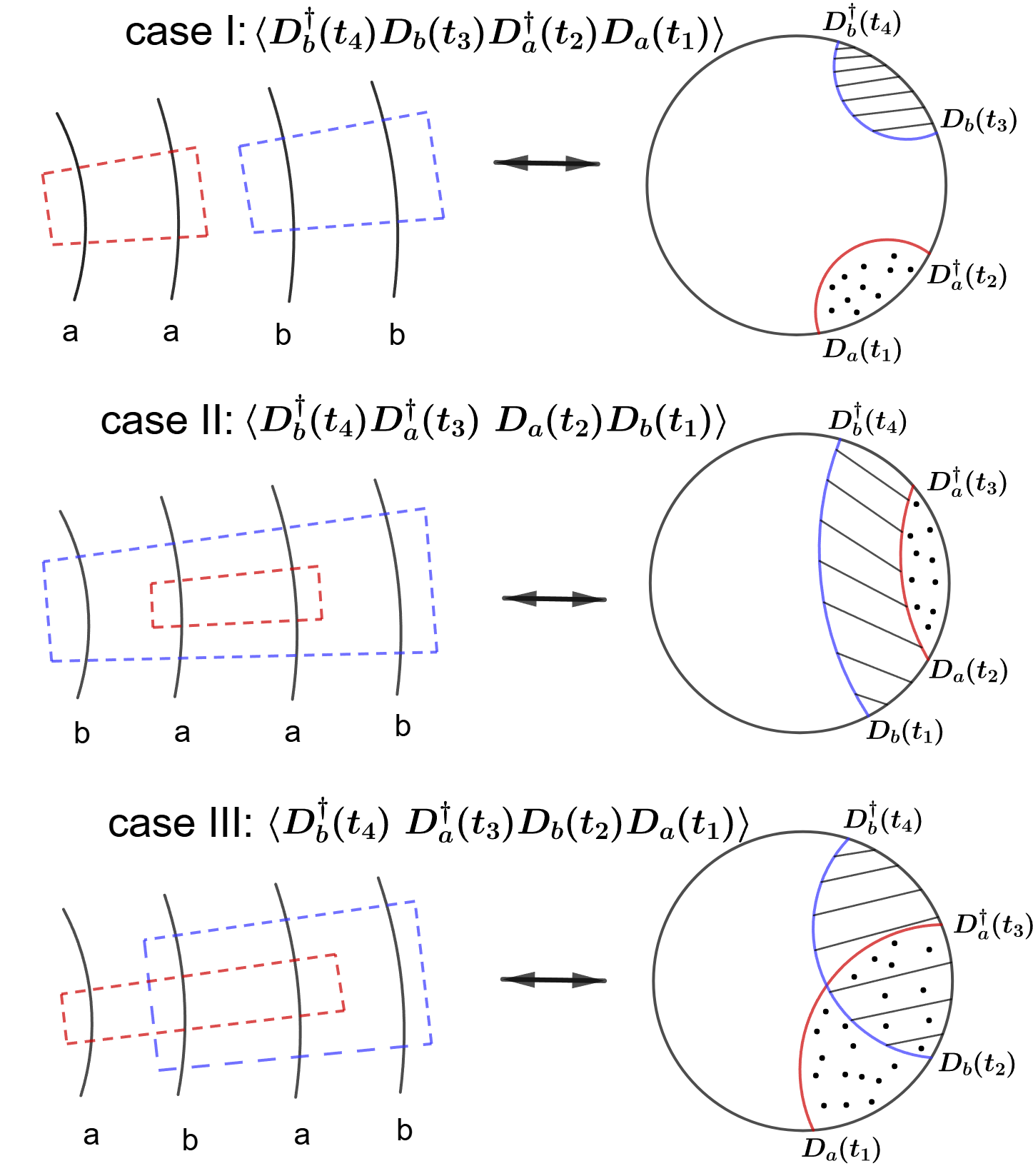}
    \caption{Three distinct orderings of four defect correlators $\langle DDDD\rangle$. Each case is shown with its  monodromy contours in field theory and the dual backreacted  geometries in gravity. Boundary monodromy contours (and the corresponding bulk domain walls) for defect $D_a$ are drawn in red, while those for defect $D_b$ are in blue.}
    \label{fig:adscft2}
\end{figure}

\subsection{ETH analysis}

We consider the following three kinds of four point functions $\langle D_b^{\dagger}(t_4) D_b(t_3) D^{\dagger}_a(t_2) D_a(t_1) \rangle$, $\langle D_b^{\dagger}(t_4) D_a^{\dagger}(t_3)  D_a(t_2) D_b(t_1)  \rangle$ and $\langle D^{\dagger}_b(t_4)  D^{\dagger}_a(t_3) D_b(t_2) D_a(t_1) \rangle$ for scalar line defects $D_a$ and $D_b$, which are ordered in the Euclidean time such that $t_4>t_3>t_2>t_1$. In this subsection, we study both the microcanonical and canonical correlators with the ETH ansatz, as we did in the two-point function case. 

By inserting  complete basis of energy eigenstates, the first microcanonical correlator 
$G_{E_H;4}^{1}(\{t_i\})$ can be expressed as  
\be\ba
   G_{E_H;4}^{1}(\{t_i\})&\equiv \langle E_H| D_b^\dagger(t_4) D_b(t_3) D^{\dagger}_a(t_2) D_a(t_1)| E_H \rangle\\
    &= \sum_{E_2, E_3, E_4} e^{ -t_{14}E_1-t_{21}E_2-t_{32}E_3-t_{43}E_4} \times \overline{D_{a,12}D^\dagger_{a,23}D_{b,34}D^\dagger_{b,41}} \ ,\label{411}
\ea\ee
where $t_{ij}=t_i-t_j$, $E_1=E_H$, and $D_{a,ij},D_{b,ij}$ are defined in \eqref{eth1} for defects $D_a$ and $D_b$. Using the ansatz \eqref{ETH}  and assuming that the Gaussian random variables $R_{ij}$  do not have non-trivial correlation between different types of defects $D_{a}$ and $D_{b}$, the average  in \eqref{411} becomes
\be  \overline{D_{a,12}D^\dagger_{a,23}D_{b,34}D^\dagger_{b,41}}=\delta_{13} \ , e^{-f_{a}(E_1,E_2)-f_{b}(E_1,E_4)}\ .  \label{4eth1}  \ee 
When we replace the summation by an integral as we have done before, the delta function should be also replaced by
\be
\delta_{ij}=\delta(E_i-E_j)e^{-S(E_j)}\ .
\ee
Consequently,  \eqref{411} can be written as
\begin{align}
    G_{E_H;4}^{1}& = \int d E_2 d E_4 \ , e^{S(E_2)+S(E_4)-(t_{14}+t_{32})E_1-t_{21}E_2-t_{43}E_4 -f_{a}(E_1,E_2)-f_{b}(E_1,E_4) } \nn \\
    & = e^{S(\tilde M_{12})+S(\tilde M_{34})-(t_{14}+t_{32})E_1-t_{21}M_{12}-t_{43}M_{34} -f_{a}(E_1,\tilde M_{12})-f_{b}(E_1,\tilde M_{34}) }\  ,
\end{align}
where we have used saddle point approximation in second line, and denoted $\tilde M_{12},\tilde M_{34}$ to be the saddle values of $E_2,E_4$ respectively which
satisfy the following equations
\be\ba\label{eq:ETH:case1}
\beta_{E_4}-t_{43}-\p_{E_4}f_b(E_1,E_4)\big|_{E_4=\tilde M_{34}}=\beta_{E_2}-t_{21}-\p_{E_2}f_a(E_1,E_2)\big|_{E_2=\tilde M_{12}}=0\ ,
\ea\ee
with $\beta_{E_i}=\p_{E_i}S(E_i)$.
Similarly, for the second correlator $G_{E_H;4}^{2}(\{t_i\})$, we have 
\begin{align}
    G_{E_H;4}^{2}(\{t_i\}) & \equiv \langle E_H| D_b^\dagger(t_4) D_a^\dagger(t_3) D_a(t_2) D_b(t_1)| E_H \rangle \nn \\
    &= \sum_{E_2,E_3,E_4} e^{ -t_{14}E_1-t_{21}E_2-t_{32}E_3-t_{43}E_4} \times \overline{\underbrace{D_{b,12}D_{a,23}D_{a,34}^\dagger D_{b,41}^\dagger}_{\propto \ , \bm{\delta_{24}} }} \nn \\
    &=\int d E_2 d E_3 \ , e^{S(E_2)+S(E_3)-t_{14} E_1-(t_{43}+t_{21}) E_2-t_{32} E_3 -f_{a}(E_1,E_2)-f_{b}(E_2,E_3) } \nn \\
    &= e^{S(\tilde M_{14})+S(\tilde M_{23})-t_{14} E_1-(t_{43}+t_{21})\tilde M_{14}-t_{32} \tilde M_{23} -f_{a}(E_1,\tilde M_{14})-f_{b}(\tilde M_{23},\tilde M_{14}) }\ ,
\end{align}
where $\tilde M_{14},\tilde M_{23}$ are the saddle values of $E_2,E_3$ respectively solved by the following equations
\begin{align}
\beta_{E_2}-(t_{43}+t_{21})-\p_{E_2}(f_a(E_1,E_2)+f_b(E_2,E_3))=\beta_{E_3}-t_{32}-\p_{E_3}f_b(E_2,E_3)=0\ .
\label{case2}
\end{align}
For the third correlator $G_{E_H;4}^{3}(\{t_i\})$, we have 
\begin{align}
    G_{E_H;4}^{3}(\{t_i\}) & = \langle E_H| D_b^\dagger(t_4) D^\dagger_a(t_3) D_b(t_2) D_a(t_1)| E_H \rangle \nn \\
    &= \sum_{E_2,E_3,E_4} e^{ - t_{14} E_1-t_{43} E_2-t_{32} E_3-t_{21} E_4} \times \overline{\underbrace{ D_{a,12} D_{b,23} D_{a,34}^{\dagger }  D_{b,41}^{\dagger }}_{\propto \ , 0 } }=0 \ ,
\end{align}
As we can see, there is no consistent way of contracting the Gaussian random variables so $G_{E_H;4}^{3}(\{t_i\})$ vanishes at the semiclassical level.

Now we come to the canonical correlators. They can be obtained from the corresponding microcanonical correlators via a Laplace transformation and the results can be easily written down
\begin{align}\label{G4:beta}
  Z_{\b} G_{\b;4}^{i}&= \int dE_He^{-\beta E_H+S(E_H)}G^i_{E_H;4} = e^{-\beta M_i+S(M_i)+\log G^i_{M_i;4}}\ ,\quad i=1,2\ ,
\end{align}
where $M_i$ is the saddle value of $E_H$ in  each case satisfying
\be\label{saddle:EH:4pt}
\beta_{E_H}+\p_{E_H}\log G^i_{E_H;4}-\beta\Big|_{E_H=M_i}=0\ .
\ee
Since we have been using the saddle point approximation, time derivatives of the correlators are simply given by
\be\label{dt:G4}
\p_{t_k}\log G^i_{E_H;4}=E_{k+1}-E_k\ ,
\ee
for $i=1,2$ and $k=1,2,3,4$ with $E_5=E_1=E_H$. Values of $E_2,E_3,E_4$ are determined by the corresponding saddle point equations in each case. Time derivatives of the canonical correlators take the same form as \eqref{dt:G4} with $E_H$ determined by \eqref{saddle:EH:4pt}.

\subsection{Vacuum Virasoro block}

The procedures for calculating  vacuum Virasoro blocks of the microcanonical correlators $G_{E_H;4}^i$ are similar to previous discussion.  The three different orders would result in three different monodromy contours, see figure \ref{fig:adscft2}. 

\begin{figure}
    \centering
    \includegraphics[width=0.7\textwidth]{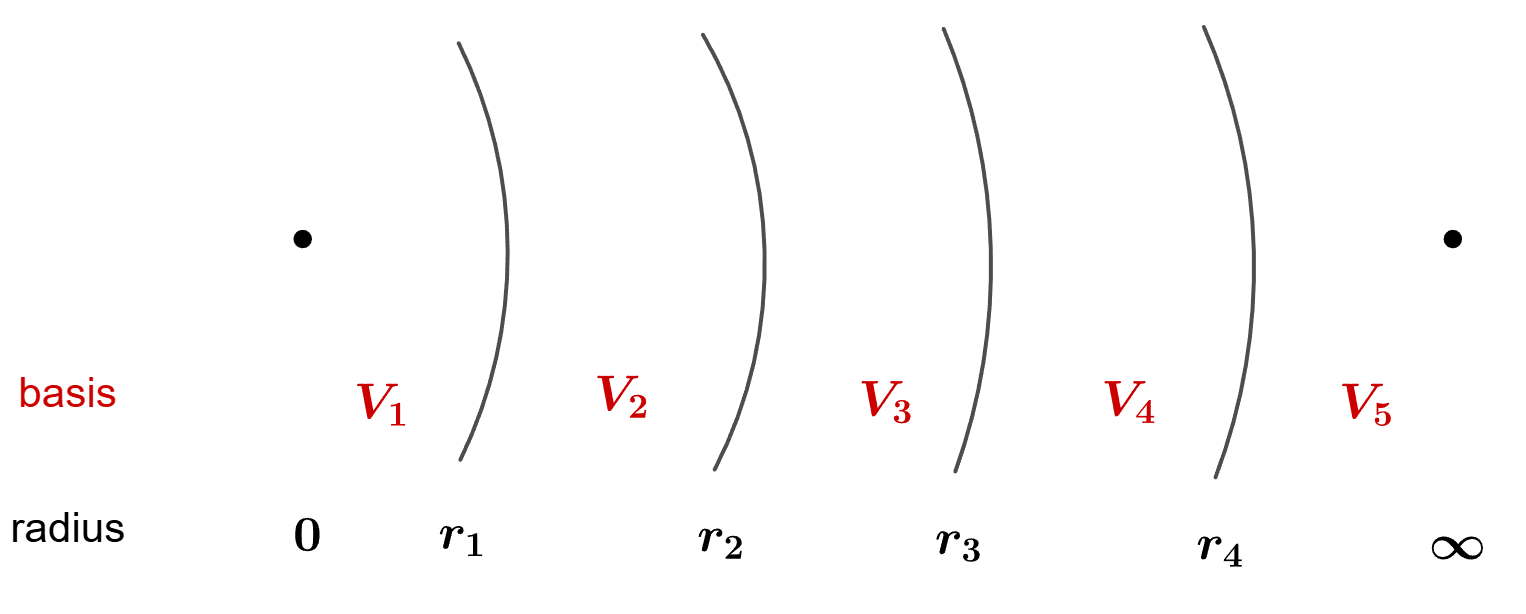}
        \caption{Concrete configuration and notations for field correlator $\langle E_H|DDDD|E_H \rangle $ with four line defects. The basis $V_i$  \eqref{bs4shell} are shown in each region.}
    \label{fig:field4setup}
\end{figure}
We begin by solving the Fuchsian equation  \eqref{diff-eq}. The stress tensor $T_4(z)$ for four defects can be computed in the continuous limit $n\to\infty$ and the result for all three cases can be written uniformly as
\be\label{STR:4pt}
 T_4  =-\sum_{k=1}^4\frac{6h_kr_k\delta(|z|-r_k)}{cz^2}+\sum_{l=1}^5\frac{1-4\rho_l^2}{4z^2}\Theta((|z|-r_{l-1})(r_{l}-|z|))\ ,
\ee
with 
\be
r_k=e^{t_k}\ ,\quad t_0=-\infty\ ,\quad t_5=\infty\ , \quad \rho_1=\rho_5=\sqrt{-\frac{3E_H}{c}}\ .
\ee
Values of $ h_k$ are assigned for different cases as follows
\be\ba
   {} & \{ {h}_{1,2}=h_{D_a}, \ {h}_{3,4}=  {h}_{D_b} \}  \quad  \longleftrightarrow \quad G_{E_H;4}^1 \ ,\\
{}    & \{  {h}_{2,3}=  {h}_{D_a}, \  {h}_{1,4}=  {h}_{D_b} \} \quad\longleftrightarrow \quad  G_{E_H;4}^2\ , \\
  {}   & \{  {h}_{1,3}=  {h}_{D_a}, \  {h}_{2,4}=  {h}_{D_b} \} \quad   \longleftrightarrow \quad  G_{E_H;4}^3\ .\label{4choice}
\ea\ee
The remaining parameters $\rho_k$ for $k=2,3,4$ in \eqref{STR:4pt} are related to the time derivatives of the correlators via
\be\label{dt:G4:mono}
\frac{d}{dt_k}\log G^i_{E_H;4}=\frac{c(\rho^2_{k}-\rho_{k+1}^2)}{3}\ ,\quad i=1,2,3\ .
\ee
They are determined by the monodromy conditions as we will specify soon.
Having obtained \eqref{STR:4pt}, the equation \eqref{diff-eq} can again be solved  and the solution is a linear combination of the following basis in different regions as illustrated in figure \ref{fig:field4setup},
\be   V_{i}=\big(z^{\frac{1}{2}-\r_i}, z^{\frac{1}{2}+\r_i} \big)^{T}\ ,   \quad r_{i-1}<|z|<r_{i} \label{bs4shell}\ . \ee 
 A consistent solution to \eqref{diff-eq} requires the basis $V_i$ 
changes to $J_iV_{i+1}$ as we cross the defect at $r_i$ with the transfer matrix $J_i$ satisfying 
\be   V_i=J_i V_{i+1}\ , \quad V_i^{'}-J_i V_{i+1}^{'}=-\frac{6h_i}{cz} V_i   \  , \quad |z|=r_i\ . \label{555} \ee 
The solutions are similar to \eqref{solJ12} and given by
\be  \label{matmat4}
J_i=\frac{1}{2\rho_{i+1}}\left(
\begin{array}{cc}
z^{-\rho_i+\rho_{i+1}}(- \frac{6h_i}{c}+\rho_i+\rho_{i+1})&z^{-\rho_i-\rho_{i+1}}( \frac{6h_i}{c}-\rho_i+\rho_{i+1})\\
z^{\rho_i+\rho_{i+1}}(- \frac{6h_i}{c}-\rho_i+\rho_{i+1})&z^{\rho_i-\rho_{i+1}}( \frac{6h_i}{c}+\rho_i+\rho_{i+1})
\end{array}
\right)\ ,\quad  |z|=r_i\ . \ee 
Now we are ready to impose monodromy conditions, where there are two independent ones for each case. For later convenience, we define
\be S^\pm_i[m,n]=\left(\frac{6h_{D_i}}{c}\pm\r_{m}+\r_{n}\right)\left(\frac{6h_{D_i}}{c}\pm\r_{m}-\r_{n}\right)\ , \quad i=a,b\ .\label{42notation}\ee 
To compute $G_{E_H;4}^1$, the monodromy conditions for the vacuum Virasoro block are 
\be  
J_1(r_1 e^{i \th_1}) J_2(r_2 e^{i \th_1}) J_2^{-1}(r_2 e^{i \th_2}) J_1^{-1}(r_1 e^{i \th_2})
=J_3(r_3 e^{i \th_3}) J_4(r_4 e^{i \th_3}) J_4^{-1}(r_4 e^{i \th_4}) J_3^{-1}(r_3 e^{i \th_4})= \textbf{1}_{2\times 2} \ .
\ee
Solving above equations giving rise to
\be \label{soshell41}
     \r_1=\r_3\ ,\quad e^{2\rho_2t_{21}}=\frac{S^-_a[2,1]}{S^-_a[2,3]}\ ,\quad e^{2\rho_4t_{43}}=\frac{S^-_b[4,3]}{S^+_b[4,1]}\ .
\ee 
Similarly, the monodromy conditions used to compute $G_{E_H;4}^2$ are given by
\be \ba  
{}&J_2(r_2 e^{i \th_1}) J_3(r_3 e^{i \th_1}) J_3^{-1}(r_3 e^{i \th_2}) J_2^{-1}(r_2 e^{i \th_2}) \nn \\
=&
   J_1(r_1 e^{i \th_3}) J_2(r_2 e^{i \th_3}) J_3(r_3 e^{i \th_3}) J_4(r_4 e^{i \th_3})J_4^{-1}(r_4 e^{i \th_4}) J_3^{-1}(r_3 e^{i \th_4}) J_2^{-1}(r_2 e^{i \th_4}) J_1^{-1}(r_1 e^{i \th_4})
   \\=& \textbf{1}_{2\times 2}\ .
    \ea
\ee
Solving above monodromy conditions, one gets
\be \label{soshell42}
\rho_2=\rho_4\ ,\quad e^{2\rho_3t_{32}}=\frac{S^-_a[3,2]}{S^+_a[3,4]}\ ,\quad e^{2\rho_4(t_{43}+t_{21})}=\frac{S^-_a[4,3]S^-_b[4,1]}{S^+_a[4,3]S^+_b[4,1]}\ .
\ee 
Up to now, we have given the monodromy equations for the first two cases. By solving either \eqref{soshell41} or \eqref{soshell42}, the parameters $\rho_{2,3,4}$ are determined and so is the time derivative of the correlator using \eqref{dt:G4}. For the third correlator $G_{E_H;4}^3$, the corresponding monodromy conditions are
\be  \label{mono:case3}
\ba
 {}& J_1(r_1 e^{i \th_1})J_2(r_2 e^{i \th_1}) J_3(r_3 e^{i \th_1}) J_3^{-1}(r_3 e^{i \th_2}) J_2^{-1}(r_2 e^{i \th_2})J_1^{-1}(r_1 e^{i \th_2}) \\
=&J_2(r_2 e^{i \th_1})J_3(r_3 e^{i \th_1}) J_4(r_4 e^{i \th_1}) J_4^{-1}(r_4 e^{i \th_2}) J_3^{-1}(r_3 e^{i \th_2})J_2^{-1}(r_2 e^{i \th_2})
 \\=& \textbf{1}_{2\times 2}\ .
\ea
\ee
Since \eqref{mono:case3} holds for any $\theta_{1,2}$, \eqref{mono:case3} is equivalent to requiring
\be
\frac{d}{d\theta}\left[J_1(r_1 e^{i \th})J_2(r_2 e^{i \th}) J_3(r_3 e^{i \th})\right]=\frac{d}{d\theta}\left[J_2(r_2 e^{i \th_1})J_3(r_3 e^{i \th_1}) J_4(r_4 e^{i \th_1}) \right]=\textbf{0}_{2\times2}\ .
\ee
The diagonal parts of above equations imply $\rho_2=\rho_4=\rho_1$. The remaining off-diagonal equations are over-constrained for $\rho_3$ and therefore no solution exists.

\subsection{Gravitational action}

\begin{figure} 
    \centering
    \includegraphics[width=0.4\linewidth]{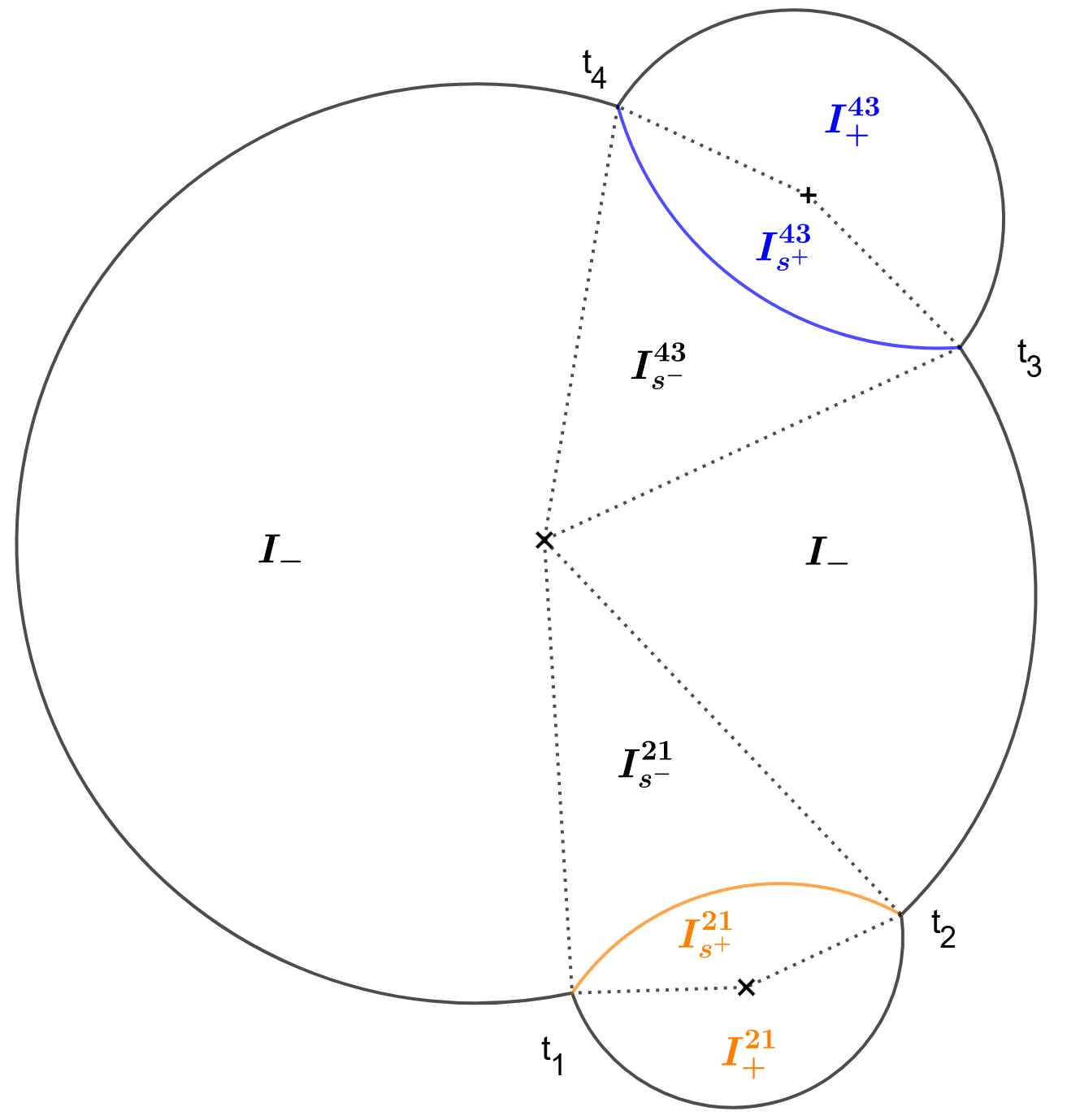}
    \hfill
\includegraphics[width=0.55\linewidth]{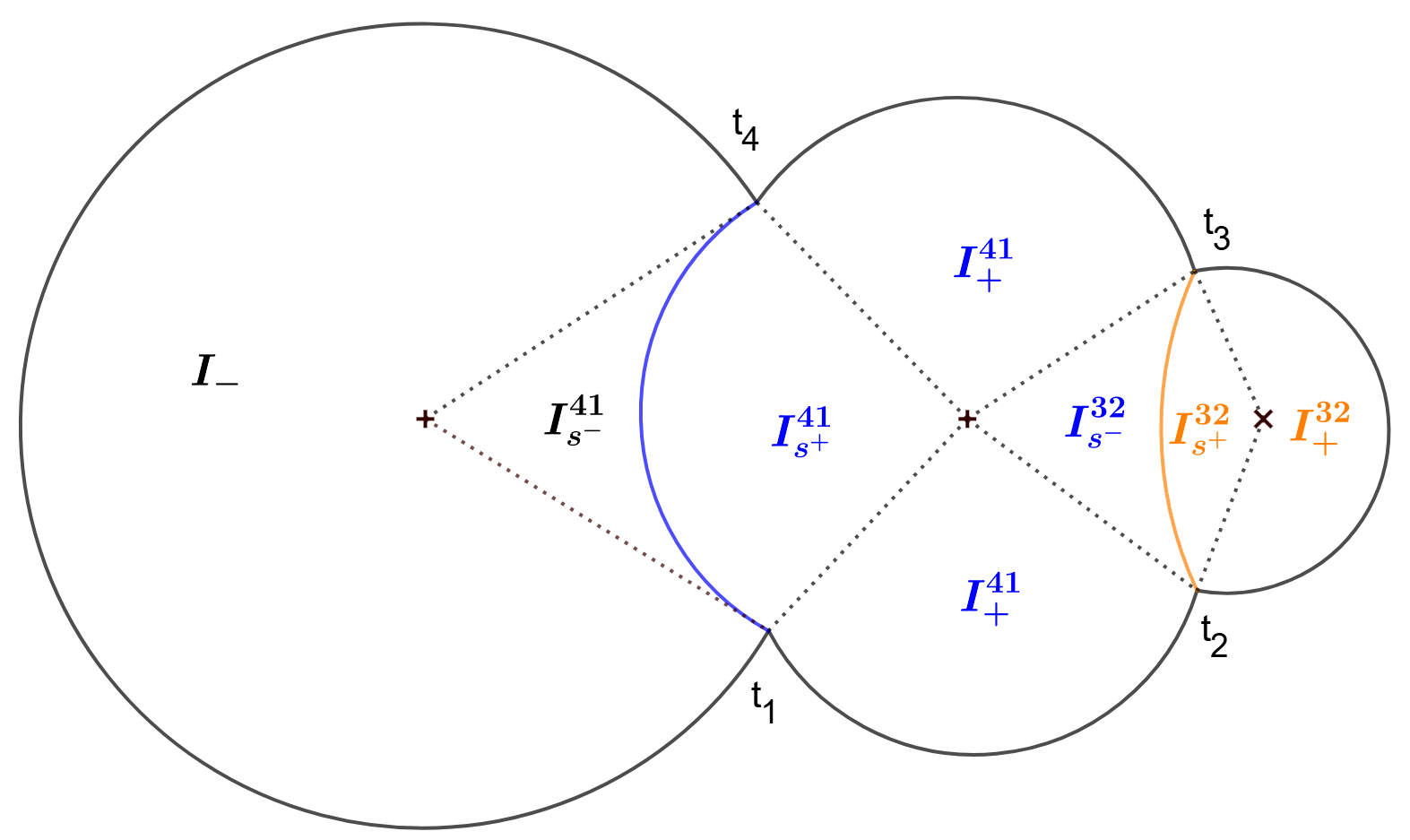}
    \caption{Concrete configurations and notations for backreacted geometries with two domain walls for equation \eqref{531} (left) and \eqref{541} (right), which correspond to case I and II in figure \ref{fig:adscft2}. Different domain walls and action terms "inside" them are shown by different colors.}
    \label{fig2}
\end{figure}
On the gravity side, to compute $G_{\beta;4}^i$, we need to solve the backreacted geometries which consist of three BTZ black holes glued together through two domain walls. Since there is no saddle geometry where the two domain walls cross each other, the third case as shown in \ref{fig:adscft2} has no stable solution. For the first two cases, computations can be performed in an analogous way to the two-defect case, and we divide the spacetime into different pieces as shown in figure \ref{fig2}. For each closed circle formed by a piece of the thermal circle as well as the domain walls, there is one saddle point equation. As a result, masses of the three black holes in each case can be determined by three saddle equations and the thermal correlators are given by evaluating the corresponding on-shell actions. 

For the first case in figure  \ref{fig2}, we denote  masses of three black holes as $M_-,M_{34}$ and $M_{12}$.
The saddle point equations for them are given by
\be\ba
    \beta_{M_-}&=\beta-(t_{41}-t_{32})+\Delta t_{-}^{34}+ \Delta t_{-}^{12}\ , \\
     \beta_{M_{34}}&=t_{43}+\Delta t_{+}^{34} \ ,\\
     \beta_{M_{12}}&=t_{21}+\Delta t_{+}^{12}\ , \label{530}
\ea\ee
where $\Delta t^{ij}_\pm$ denotes the time elapsed by the domain wall which travels from $t_i$ to $t_j$ measured in the corresponding region. 
The total on-shell gravitational action is given by
\be I= I_{-}+I_{+}^{43}+I_{+}^{21}+I_{s^{-}}^{43}+ I_{s^{+}}^{43}+I_{s^{-}}^{21}+ I_{s^{+}}^{21}+I_{\text{ct}}\ , \label{531} \ee
where  
\be\ba\label{acts1}
   & I_{\text{ct}}=-\frac{m_a+m_b}{2G}\log r_\infty\ ,  \\
  &  I_+^{43}= -\frac{t_{43} M_{34}}{8G}, \quad I_+^{21}= -\frac{t_{21} M_{12} }{8G},\quad I_- =-\frac{(\b-t_{41}+t_{32}) M_- }{8G}\ ,\\
  &I^{43}_{s^+}+I^{43}_{s^-}=\frac{m_b}{2G}\cosh^{-1}\frac{r_\infty}{r_{b*}},\quad I^{21}_{s^+}+I^{21}_{s^-}=\frac{m_a}{2G}\cosh^{-1}\frac{r_\infty}{r_{a*}}\ ,
\ea\ee
with
\begin{align}
    r_{a*}=r_{-}^2+\big( \frac{M_{-}-M_{21}}{2m_a}-\frac{m_a }{2}\big)^2\ , \ r_{b*}=r_{-}^2+\big( \frac{M_{-}-M_{43}}{2m_b}-\frac{ m_b}{2} \big)^2\ .
\end{align}
Similarly, saddle equations that determine the masses of black holes in the second case are given by
\be\ba
    &   \beta_{M_-}=\beta-\tau_{41}+\Delta t_{-}^{41}\  , \\
    &   \beta_{M_{41}}=\tau_{43}+\tau_{21}+\Delta t_{+}^{41}+\Delta t_{-}^{32}\ ,\\
    & \beta_{M_{32}}=\tau_{32}+\Delta t_{+}^{32}\ . \label{540}
\ea\ee
And the on-shell gravitational action is
\be I= I_{-}+I_{+}^{41}+I_{+}^{32}+I_{s^{-}}^{41}+ I_{s^{+}}^{41}+I_{s^{-}}^{32}+ I_{s^{+}}^{32}+I_{\text{ct}}\ , \label{541}\ee
where each term can be  calculated similarly to \eqref{acts1} and we will not give their explict forms here for simplicity.

\subsection{Matching}
Using the choice of the envelop function \eqref{envelop}, it is easy to show that saddle equations \eqref{eq:ETH:case1} match with monodromy equations \eqref{soshell41} under the identification
\be
\tilde M_{12}=-\frac{c\rho_2^2}{3}\ ,\quad\tilde  M_{34}=-\frac{c\rho_4^2}{3}\ .
\ee 
As a result, ETH analysis and vacuum Virasoro block give the same microcanonical four point defect function $G_{E_H;4}^1$ by virtue of \eqref{dt:G4} and \eqref{dt:G4:mono}. Similarly, comparing \eqref{eq:ETH:case1}, \eqref{saddle:EH:4pt} with gravitational saddle equations \eqref{530} and using the dictionaries $m_a=8Gh_{D_a},m_b=8Gh_{D_b}$, we have the identifications 
\be
\tilde M_{12}=  M_{12}\ ,\quad \tilde M_{34}=  M_{34}\ ,\quad M_1=M_-\ .
\ee
This also suggests that the thermal four point function agrees with each other by directly comparing \eqref{G4:beta} with \eqref{531}.
Matching for the second case among three methods can be done in exactly the same way.

\subsection{Higher point generalizations}
\begin{figure} 
    \centering
    \includegraphics[width=0.35\linewidth]{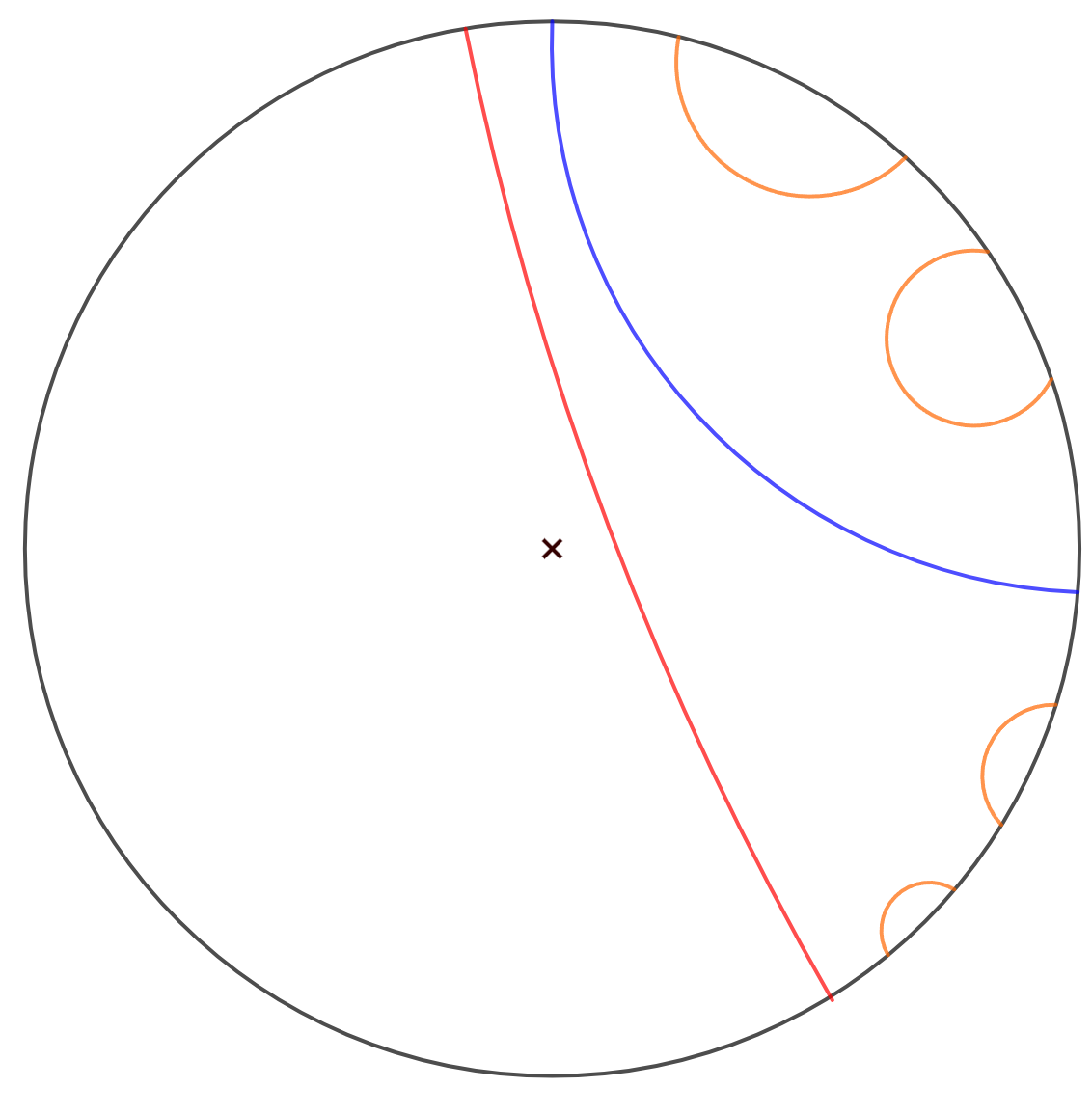}
    \hfill
\includegraphics[width=0.6\linewidth]{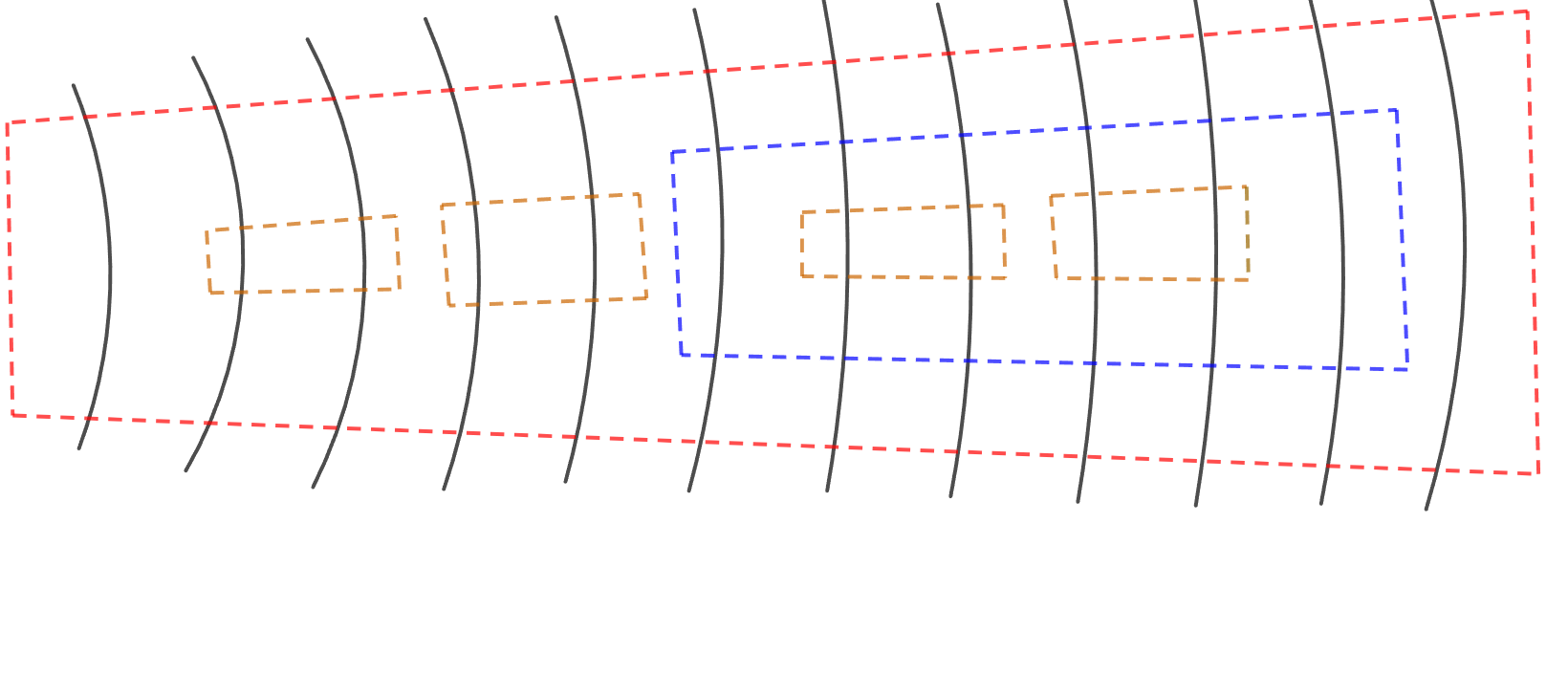}
    \caption{An multi-defect example showing the match between gravitational configuration and monodromy contours.}
    \label{multiple}
\end{figure}
We end this section by providing a generic picture for computing  $2n$-point functions.  
More precisely,
we consider a CFT$_2$ defined on the Euclidean cylinder and insert $n$ pairs of defect operators $D,D^\dagger$ which are ordered in the Euclidean time $t_{i},i=1,\cdots 2n$ with $t_{2n}>t_{2n-1}>\cdots t_{1}$,
\be
G_{E_H;2n}(\{t_i\})\equiv\langle E_H|\prod_{i=1}^{2n} D^{\alpha_i}(t_i)|E_H\rangle\ ,
\ee
with $D^{\alpha_i}=D$ or $D^\dagger$.
The correlation functions depend on the configurations of defect $\{D^{\alpha_i}\}$ insertions. We will be interested in the vacuum Virasoro block with channel $D^\dagger D\to\ id$, therefore will have $n$ loops leading to $n$ monodromy equations.
On the gravity side, we need to consider correlation function
\be
G_{\beta;2n}(\{t_i\})\equiv\langle \prod_{i=1}^{2n} D^{\alpha_i}(t_i)\rangle_\beta\ .
\ee
This quantity also should be computed from the on-shell action of the backreacted saddle geometry. There are $n$ trajectories of dust particles propagating in spacetime which start from the boundary at insertion location of ${D}$ and end on the boundary at insertion location of ${D}^\dagger$.
Since there is no stable saddle geometry  where the two domain wall cross with each other, the $n$ trajectories divide the spacetime into $n+1$ regions with each being locally a BTZ black hole. The masses of black holes are determined by $n+1$ gravitational boundary conditions obtained by properly identifying the thermal circles with the elapsed times and boundary insertion times.
From the perspective of ETH ansatz, the microcanonical correlation function can be written as
\be
 G_{E_H;2n}(\{t_i\})=\int\prod_{i=2}^{2n} dE_ie^{-\mathcal{I}_{2n}}\overline{\prod_{i=1}^{2n} R^{\alpha_i}_{i,i+1}}\ ,
\ee
with $R_{i,i+1}^{\alpha_i}=R_{i,i+1}$ or $R^*_{i,i+1}$ being Gaussian random variables and 
\be
\mathcal{I}_{2n}=\sum_{i=2}^{2n}S(E_i)+t_i(E_{i+1}-E_i)-f(E_{i},E_{i+1}),\quad E_{i+1}=E_H\ .
\ee
We only consider the leading contribution from Gaussian statistics in the semiclassical limit, which constraints the way of contracting the random variables. 

So far, we have described three perspectives of computing higher point correlation functions with multiple defects. We conjecture that the following procedures are equivalent:
\begin{itemize}
    \item On the field theory side, imposing trivial monodromy condition on a loop crossing a pair of defects (located at two ends of the monodromy contour) ${D}^\dagger(t_i)$ and ${D}(t_j)$, or performing an OPE ${D}^\dagger(t_i){D}(t_j)\to id$, see right figure in figure \ref{multiple}.
    \item On the gravity side, having a trajectory of dust particles traveling between $t=t_i$ and $t=t_j$, see left figure in figure \ref{multiple}.
    \item In ETH analysis, contracting the Gaussian random variables $R_{i,i+1}$ with $R^*_{j,j+1}$ giving rise to $\delta_{i,j+1}\delta_{i+1,j}$.
\end{itemize}

\section{Outlook}
In this section, we list some directions that can be explored in the future.

\subsection{Non-vacuum block}

\begin{figure}
    \centering
    \includegraphics[width=0.88\linewidth]{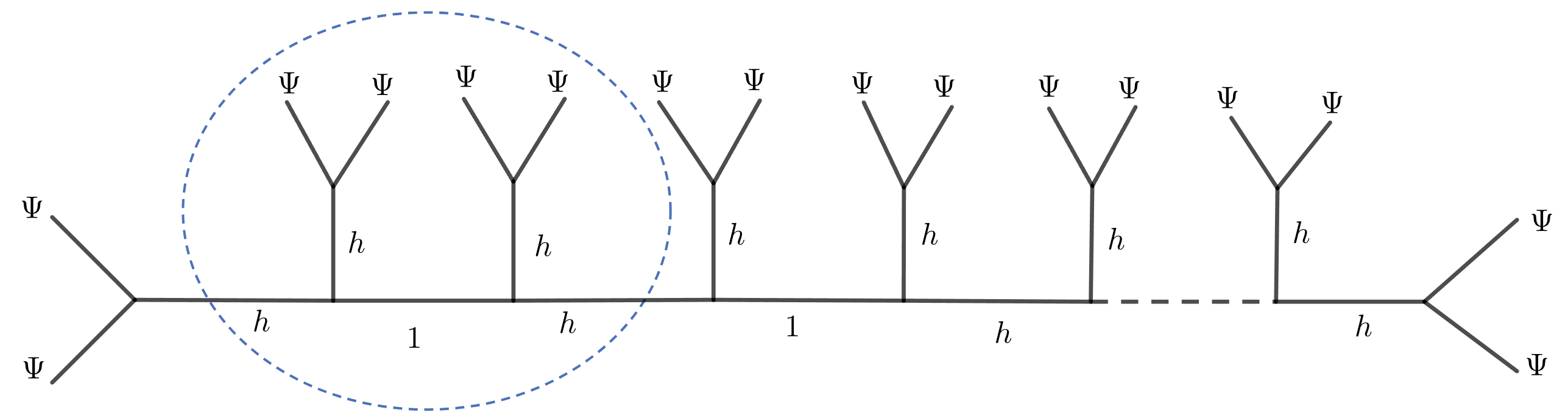}
    \caption{OPE channels for two "discrete" shell operators. If each group of four local insertions is treated as a single unit, the OPE channel exhibits a full rotational symmetry.}
    \label{fig:channel}
\end{figure}

The successful computation of the vacuum Virasoro conformal block relies on the cylindrical rotational symmetry in both the defect configurations and  the choice of internal OPE channels. More specifically, every internal channel between one pair of local operators constituting the two defects has been taken to be the identity channel, which trivially respects the rotational invariance. Actually, rotational symmetry can also be preserved if we group every  four local operators together with  a specific OPE and treat such group as the building block in constructing the entire OPE channel, see figure \ref{fig:channel}. Studying the monodromy problem of this configuration could provide analytic control over  non-vacuum blocks. {This advance is possible to pave the way toward a field theoretic derivation of the island formula, which is observed in late time black hole evaporation related to the phase transition between two Euclidean saddles. Capturing the emergence of island on the CFT side requires two competing contributions, making nonvacuum conformal blocks indispensable. }

\subsection{Analytic continuation}
Another important issue concerns the behavior of the microcanonical correlator in the late Lorentzian time regime. For local operators, the appearance of forbidden singularities in the microcanonical correlator in the probe limit implies an exponential decay in the Lorentzian regime, giving rise to the information loss problem \cite{Fitzpatrick:2016ive}. The resolution to these forbidden singularities or the exponential decay of the Virasoro block has been extensively studied \cite{Fitzpatrick:2016mjq,Chen:2017yze}. In our research, we have been focusing on the Euclidean regime.  It would be interesting to study the solution to the monodromy equation and its implications on the correlator in Lorentzian time. The holographic description for the real time correlator of local operators can also be found in \cite{Skenderis:2008dh, Skenderis:2008dg}.
Another interesting object that can be studied with the knowledge of analytic continuation would be the out of time correlator(OTOC) \cite{Dolan:2000ut}. For local operators, the OTOC plays an important role in quantifying the chaotic property of the system \cite{Larkin:1964wok,Roberts:2014ifa,Rozenbaum:2016mmv}. In the context of AdS/CFT correspondence, the OTOC exhibits an exponential decay saturating the Lyapunov  bound at late Lorentzian time and shows that black hole is  maximally chaotic \cite{Shenker:2013pqa,Shenker:2013yza,Shenker:2014cwa}.  It would be also interesting to study the OTOC for nonlocal defect operators.

\subsection{Defect RG flow}

Recently, the studies of the RG  flow of defects in \cite{Kobayashi:2018lil, Cuomo:2021rkm} have suggested that a special kind of codimension one generalized pinning field defect
\be D_h(O)=\left[e^{h\int_{\Sigma} O(x)} \right]_{ren} \ee 
would flow to a nontrivial conformal defect in the IR and exhibit a factorization property \cite{Popov:2025cha}. The non-conformal line defect studied here is closely analogous to this pinning field defect in two dimensions. Explicit perturbative results for two-defect correlators, valid over a wide range of parameters, have been worked out in \cite{Chen:2024hqu}. {It is therefore natural to ask whether, under defect RG flow, the line defect we studied likewise flows to a conformal defect that factorizes into conformal boundary conditions for two disconnected regions of
the spacetime.} One may address this question either via boundary RG techniques or by constructing a bulk holographic RG flow. Along the way,  a clearer understanding of the continuous limit of infinite operator products and the various approximation schemes in computing these correlators is needed.

\section*{Acknowledgement}
We would like to thank Yuya Kusuki, Joan Sim\'on and Jie-qiang Wu for helpful discussions. YL also thanks
Kyushu University Institute for Advanced Study, where the conference “Kyushu
IAS-iTHEMS workshop: Non-perturbative methods in QFT” was helpful in completing this work.

\appendix
\section{Junction Condition with Spinning Sources}\label{Appendix}
In this appendix, we briefly review \cite{Giacomini:2006um} for the junction condition in the presence of a spinning domain wall and apply their result to our calculations.  In the first order formalism, 
the Lagrangian for matter field coupled to  Einstein gravity with negative cosmological constant  is given by
\be
\mathcal{L}=(R^{ab}\wedge \mathrm{e}^c+\frac{1}{3}\mathrm e^a\wedge \mathrm e^b\wedge \mathrm e^c)\epsilon_{abc}+\mathcal{L}_{matter}\ ,
\ee
where $\mathrm{e}^a=\mathrm{e}^a_\mu dx^\mu$ is the vielbein one-form, $\omega^{ab}$ is the spin connection and $R^{ab}=d\omega^{ab}+\omega^a_c\wedge \omega^{cb}$ is the curvature two-form. Varying the Lagrangian with respect to the vielbein and spin connection respectively gives rise to
\be\ba\label{E-L:eq}
{}&R^{ab}\wedge dx^\mu\epsilon_{abc}=-2\sqrt{g}\mathcal T^\mu_c d^3x\ ,
\\{}&dx^\mu\wedge T^c\epsilon_{abc}=\sqrt{g}\mathcal S^\mu_{ab}d^3x\ ,
\ea
\ee
where $T^a=d\omega^a+\omega^a_b\wedge e^b$ is the torsion, $\mathcal T^\mu_a$ is the stress-energy tensor and $\mathcal S^\mu_{ab}$ is the spin current. Given that $\mathcal T^\mu_a$ and $\mathcal S^\mu_{ab}$ localize at the worldvolume $\mathcal{W}$ of the domain wall, we can define the localized quantities $T^\mu_a$ and $S^\mu_{ab}$ as
\be
\mathcal T^\mu_a=T^\mu_a\delta(\mathcal{W})\ ,\quad \mathcal S^\mu_{ab}=S^\mu_{ab}\delta(\mathcal{W})\ .
\ee
Let $\mathcal{D}$ be a bulk region containing $\mathcal{W}$ with arbitrarily small width.  Integrating the right hand side of \eqref{E-L:eq} over $\mathcal{D}$ gives
\be\label{integrate-RHS}
-2\int_{\mathcal{D}}d^3x\sqrt{g}\mathcal{T}^\mu_a=-2\int_{\mathcal{W}}\sqrt{h}T^\mu_a\ ,\quad \int_{\mathcal{D}}d^3x\sqrt{g}\mathcal{S}^\mu_{ab}=-2\int_{\mathcal{W}}\sqrt{h}S^\mu_{ab}\ ,
\ee
where $h$ is the determinant of the induced metric on $\mathcal{W}$.
Similarly, integrating the left hand side of \eqref{E-L:eq} with the requirement that the vielbein and spin connection have a bounded discontinuity at $\mathcal{W}$ gives
\be\ba\label{integrate-LHS}
{}&\int_{\mathcal{D}}R^{ab}\wedge dx^\mu\epsilon_{abc}=\int_{\mathcal{W}}\iota([\omega^{ab}]\wedge dx^\mu\epsilon_{abc})\ ,
\\{}&\int_{\mathcal{D}} dx^\mu\wedge T^c\epsilon_{abc}=\int_{\mathcal{W}}\iota(dx^\mu\wedge[e^c]\epsilon_{abc})\ ,
\ea
\ee
where $\iota$ is the pull-back of diﬀerential forms to $\mathcal{W}$ and $[X]$ is defined as the discontinuity of $X$ across $\mathcal{W}$.
Equating \eqref{integrate-LHS} with \eqref{integrate-RHS} gives the junction condition in the presence of spinning domain wall which is
\be\ba\label{1st-junction}
&\iota[\omega^{ab}\wedge dx^\mu\epsilon_{abc}]=-2T^\mu_c\text{Vol}\ ,
\\&\iota[dx^\mu\wedge \mathrm e^c\epsilon_{abc}]= S^\mu_{ab}\text{Vol}\ ,
\ea\ee
with $\text{Vol}=\sqrt{h}d^2y$ being the volume form of $\mathcal{W}$.
Note that in order for \eqref{integrate-RHS} to be well defined, the volume form needs to be single valued on $\mathcal{W}$. As a result, $h$ should be continuous across the shell, i.e. $[\text{Vol}]=0$. If we further require the stress tensor and spin current to be single valued when contracted with the vielbein, it turns out that  an additional condition is required, which is
\be\label{condition2}
\iota[e^a\wedge [e^b]]=0\ .
\ee
In summary, to have a well defined junction condition from which the stress tenson and spin current can be defined unambiguously, we need to require that the induced volume form to be continuous together with \eqref{condition2}. As a result, the localized stress tensor and spin current can be computed using \eqref{1st-junction}.

Having reviewed the derivation of the junction condition, we use rotating BTZ black hole as an example to see how \eqref{condition2} are satisfied and compute the localized stress tensor and spin current as well.
Given the metric of rotating BTZ black hole \eqref{rotatingbtz}, we choose the 3d veilbeins to be
\be
\mathrm e^1=edt+jd\phi+\frac{v^r}{f(r)}dr\ ,\quad \mathrm{e}^2=\frac{r}{\tilde r}(v^rdt-v^tdr)\ ,\quad \mathrm{e}^3=\tilde r(d\phi-\frac{iJ}{2\tilde r^2}dt)-\frac{j(v^r/f(r)dr+edt)}{\tilde r}\ ,
\ee
where $\tilde r=\sqrt{r^2-j^2}$ and $v^r=r'(\ell)$ given by \eqref{geo}. 
Such choice features the advantage that the induced veilbeins take very simple form
\be
\iota\mathrm{e}^1\equiv e^1=d\ell+jd\psi\ ,\quad \iota\mathrm{e}^2=0\ ,\quad 
\iota\mathrm{e}^3\equiv e^2=\tilde rd\psi\ .
\ee
Therefore, it is straightforward to see that \eqref{condition2} is satisfied.
Define $\omega_a\equiv\omega^{ab}\epsilon_{abc}$ and we find
\be
\ba
\iota\omega_1=2ed\psi+\frac{iJ+2ej}{\tilde r^2}d\ell\ ,\quad \iota\omega_2=\frac{2rv^r}{\tilde r}d\psi\ ,\quad \iota\omega_3=-\frac{iJ+2ej}{\tilde r}d\psi\ ,
\ea
\ee
Using \eqref{1st-junction}, we can compute the stress tensor in local frames and we find the only non-vanishing component is given by
\be
T^\ell_1=\frac{[e]}{\tilde r}\ .
\ee
Similarly, we find that the only non-vanishing component of the spin current is given by
\be
S^\ell_{23}=-\frac{[j]}{\tilde r}\ .
\ee
With such results, $m$ and $s$ can indeed be regarded as the total mass and spin carried by the dust as measured by an asymptotic observer.

\bibliographystyle{fullsort}
\newpage
\bibliography{ref}

\providecommand{\href}[2]{#2}\begingroup\raggedright\begin{thebibliography}{10}

\bibitem{Maldacena:1997re}
J.~M. Maldacena, ``{The Large N limit of superconformal field theories and supergravity},'' {\em Adv. Theor. Math. Phys.} {\bf 2} (1998) 231--252, \href{http://www.arXiv.org/abs/hep-th/9711200}{{\tt hep-th/9711200}}.

\bibitem{Witten:1998qj}
E.~Witten, ``{Anti-de Sitter space and holography},'' {\em Adv. Theor. Math. Phys.} {\bf 2} (1998) 253--291, \href{http://www.arXiv.org/abs/hep-th/9802150}{{\tt hep-th/9802150}}.

\bibitem{Gubser:1998bc}
S.~S. Gubser, I.~R. Klebanov, and A.~M. Polyakov, ``{Gauge theory correlators from noncritical string theory},'' {\em Phys. Lett. B} {\bf 428} (1998) 105--114, \href{http://www.arXiv.org/abs/hep-th/9802109}{{\tt hep-th/9802109}}.

\bibitem{Saad:2019lba}
P.~Saad, S.~H. Shenker, and D.~Stanford, ``{JT gravity as a matrix integral},'' \href{http://www.arXiv.org/abs/1903.11115}{{\tt 1903.11115}}.

\bibitem{Penington:2019kki}
G.~Penington, S.~H. Shenker, D.~Stanford, and Z.~Yang, ``{Replica wormholes and the black hole interior},'' {\em JHEP} {\bf 03} (2022) 205, \href{http://www.arXiv.org/abs/1911.11977}{{\tt 1911.11977}}.

\bibitem{Almheiri:2019qdq}
A.~Almheiri, T.~Hartman, J.~Maldacena, E.~Shaghoulian, and A.~Tajdini, ``{Replica Wormholes and the Entropy of Hawking Radiation},'' {\em JHEP} {\bf 05} (2020) 013, \href{http://www.arXiv.org/abs/1911.12333}{{\tt 1911.12333}}.

\bibitem{Belin:2020hea}
A.~Belin and J.~de~Boer, ``{Random statistics of OPE coefficients and Euclidean wormholes},'' {\em Class. Quant. Grav.} {\bf 38} (2021), no.~16, 164001, \href{http://www.arXiv.org/abs/2006.05499}{{\tt 2006.05499}}.

\bibitem{Cotler:2021cqa}
J.~Cotler and K.~Jensen, ``{Wormholes and black hole microstates in AdS/CFT},'' {\em JHEP} {\bf 09} (2021) 001, \href{http://www.arXiv.org/abs/2104.00601}{{\tt 2104.00601}}.

\bibitem{Chandra:2022bqq}
J.~Chandra, S.~Collier, T.~Hartman, and A.~Maloney, ``{Semiclassical 3D gravity as an average of large-c CFTs},'' {\em JHEP} {\bf 12} (2022) 069, \href{http://www.arXiv.org/abs/2203.06511}{{\tt 2203.06511}}.

\bibitem{Strominger:1996sh}
A.~Strominger and C.~Vafa, ``{Microscopic origin of the Bekenstein-Hawking entropy},'' {\em Phys. Lett. B} {\bf 379} (1996) 99--104, \href{http://www.arXiv.org/abs/hep-th/9601029}{{\tt hep-th/9601029}}.

\bibitem{Mathur:2005zp}
S.~D. Mathur, ``{The Fuzzball proposal for black holes: An Elementary review},'' {\em Fortsch. Phys.} {\bf 53} (2005) 793--827, \href{http://www.arXiv.org/abs/hep-th/0502050}{{\tt hep-th/0502050}}.

\bibitem{Altland:2021rqn}
A.~Altland, D.~Bagrets, P.~Nayak, J.~Sonner, and M.~Vielma, ``{From operator statistics to wormholes},'' {\em Phys. Rev. Res.} {\bf 3} (2021), no.~3, 033259, \href{http://www.arXiv.org/abs/2105.12129}{{\tt 2105.12129}}.

\bibitem{DiUbaldo:2023qli}
G.~Di~Ubaldo and E.~Perlmutter, ``{AdS$_{3}$/RMT$_{2}$ duality},'' {\em JHEP} {\bf 12} (2023) 179, \href{http://www.arXiv.org/abs/2307.03707}{{\tt 2307.03707}}.

\bibitem{Anous:2021caj}
T.~Anous, A.~Belin, J.~de~Boer, and D.~Liska, ``{OPE statistics from higher-point crossing},'' {\em JHEP} {\bf 06} (2022) 102, \href{http://www.arXiv.org/abs/2112.09143}{{\tt 2112.09143}}.

\bibitem{Chen:2024hqu}
B.~Chen, Y.~Liu, and B.~Yu, ``{Correlation function of thin-shell operators},'' {\em JHEP} {\bf 08} (2024) 082, \href{http://www.arXiv.org/abs/2404.11423}{{\tt 2404.11423}}.

\bibitem{Chandra:2024vhm}
J.~Chandra, T.~Hartman, and V.~Meruliya, ``{Statistics of three-dimensional black holes from Liouville line defects},'' \href{http://www.arXiv.org/abs/2404.15183}{{\tt 2404.15183}}.

\bibitem{deBoer:2024mqg}
J.~de~Boer, D.~Liska, and B.~Post, ``{Multiboundary wormholes and OPE statistics},'' {\em JHEP} {\bf 10} (2024) 207, \href{http://www.arXiv.org/abs/2405.13111}{{\tt 2405.13111}}.

\bibitem{Schlenker:2022dyo}
J.-M. Schlenker and E.~Witten, ``{No ensemble averaging below the black hole threshold},'' {\em JHEP} {\bf 07} (2022) 143, \href{http://www.arXiv.org/abs/2202.01372}{{\tt 2202.01372}}.

\bibitem{Cotler:2022rud}
J.~Cotler and K.~Jensen, ``{A precision test of averaging in AdS/CFT},'' {\em JHEP} {\bf 11} (2022) 070, \href{http://www.arXiv.org/abs/2205.12968}{{\tt 2205.12968}}.

\bibitem{Karlsson:2021duj}
R.~Karlsson, A.~Parnachev, and P.~Tadi\'c, ``{Thermalization in large-N CFTs},'' {\em JHEP} {\bf 09} (2021) 205, \href{http://www.arXiv.org/abs/2102.04953}{{\tt 2102.04953}}.

\bibitem{Srednicki:1994mfb}
M.~Srednicki, ``{Chaos and Quantum Thermalization},'' {\em Phys. Rev. E} {\bf 50} (3, 1994) \href{http://www.arXiv.org/abs/cond-mat/9403051}{{\tt cond-mat/9403051}}.

\bibitem{Deutsch:1991msp}
J.~M. Deutsch, ``{Quantum statistical mechanics in a closed system},'' {\em Phys. Rev. A} {\bf 43} (1991), no.~4, 2046.

\bibitem{Rigol:2007juv}
M.~Rigol, V.~Dunjko, and M.~Olshanii, ``{Thermalization and its mechanism for generic isolated quantum systems},'' {\em Nature} {\bf 452} (2008), no.~7189, 854--858, \href{http://www.arXiv.org/abs/0708.1324}{{\tt 0708.1324}}.

\bibitem{DAlessio:2015qtq}
L.~D'Alessio, Y.~Kafri, A.~Polkovnikov, and M.~Rigol, ``{From quantum chaos and eigenstate thermalization to statistical mechanics and thermodynamics},'' {\em Adv. Phys.} {\bf 65} (2016), no.~3, 239--362, \href{http://www.arXiv.org/abs/1509.06411}{{\tt 1509.06411}}.

\bibitem{deBoer:2023vsm}
J.~de~Boer, D.~Liska, B.~Post, and M.~Sasieta, ``{A principle of maximum ignorance for semiclassical gravity},'' {\em JHEP} {\bf 2024} (2024) 003, \href{http://www.arXiv.org/abs/2311.08132}{{\tt 2311.08132}}.

\bibitem{Maldacena:2015waa}
J.~Maldacena, S.~H. Shenker, and D.~Stanford, ``{A bound on chaos},'' {\em JHEP} {\bf 08} (2016) 106, \href{http://www.arXiv.org/abs/1503.01409}{{\tt 1503.01409}}.

\bibitem{Fitzpatrick:2015zha}
A.~L. Fitzpatrick, J.~Kaplan, and M.~T. Walters, ``{Virasoro Conformal Blocks and Thermality from Classical Background Fields},'' {\em JHEP} {\bf 11} (2015) 200, \href{http://www.arXiv.org/abs/1501.05315}{{\tt 1501.05315}}.

\bibitem{Cotler:2016fpe}
J.~S. Cotler, G.~Gur-Ari, M.~Hanada, J.~Polchinski, P.~Saad, S.~H. Shenker, D.~Stanford, A.~Streicher, and M.~Tezuka, ``{Black Holes and Random Matrices},'' {\em JHEP} {\bf 05} (2017) 118, \href{http://www.arXiv.org/abs/1611.04650}{{\tt 1611.04650}}. [Erratum: JHEP 09, 002 (2018)].

\bibitem{Saad:2018bqo}
P.~Saad, S.~H. Shenker, and D.~Stanford, ``{A semiclassical ramp in SYK and in gravity},'' \href{http://www.arXiv.org/abs/1806.06840}{{\tt 1806.06840}}.

\bibitem{Cotler:2020ugk}
J.~Cotler and K.~Jensen, ``{AdS$_{3}$ gravity and random CFT},'' {\em JHEP} {\bf 04} (2021) 033, \href{http://www.arXiv.org/abs/2006.08648}{{\tt 2006.08648}}.

\bibitem{Altland:2022xqx}
A.~Altland, B.~Post, J.~Sonner, J.~van~der Heijden, and E.~P. Verlinde, ``{Quantum chaos in 2D gravity},'' {\em SciPost Phys.} {\bf 15} (2023), no.~2, 064, \href{http://www.arXiv.org/abs/2204.07583}{{\tt 2204.07583}}.

\bibitem{Saad:2022kfe}
P.~Saad, D.~Stanford, Z.~Yang, and S.~Yao, ``{A convergent genus expansion for the plateau},'' {\em JHEP} {\bf 09} (2024) 033, \href{http://www.arXiv.org/abs/2210.11565}{{\tt 2210.11565}}.

\bibitem{Andrei:2018die}
N.~Andrei {\em et al.}, ``{Boundary and Defect CFT: Open Problems and Applications},'' {\em J. Phys. A} {\bf 53} (2020), no.~45, 453002, \href{http://www.arXiv.org/abs/1810.05697}{{\tt 1810.05697}}.

\bibitem{Gaiotto:2014kfa}
D.~Gaiotto, A.~Kapustin, N.~Seiberg, and B.~Willett, ``{Generalized Global Symmetries},'' {\em JHEP} {\bf 02} (2015) 172, \href{http://www.arXiv.org/abs/1412.5148}{{\tt 1412.5148}}.

\bibitem{Bhardwaj:2017xup}
L.~Bhardwaj and Y.~Tachikawa, ``{On finite symmetries and their gauging in two dimensions},'' {\em JHEP} {\bf 03} (2018) 189, \href{http://www.arXiv.org/abs/1704.02330}{{\tt 1704.02330}}.

\bibitem{Chang:2018iay}
C.-M. Chang, Y.-H. Lin, S.-H. Shao, Y.~Wang, and X.~Yin, ``{Topological Defect Lines and Renormalization Group Flows in Two Dimensions},'' {\em JHEP} {\bf 01} (2019) 026, \href{http://www.arXiv.org/abs/1802.04445}{{\tt 1802.04445}}.

\bibitem{Calabrese:2004eu}
P.~Calabrese and J.~L. Cardy, ``{Entanglement entropy and quantum field theory},'' {\em J. Stat. Mech.} {\bf 0406} (2004) P06002, \href{http://www.arXiv.org/abs/hep-th/0405152}{{\tt hep-th/0405152}}.

\bibitem{Cardy:2013nua}
J.~Cardy, ``{Some results on the mutual information of disjoint regions in higher dimensions},'' {\em J. Phys. A} {\bf 46} (2013) 285402, \href{http://www.arXiv.org/abs/1304.7985}{{\tt 1304.7985}}.

\bibitem{Bianchi:2015liz}
L.~Bianchi, M.~Meineri, R.~C. Myers, and M.~Smolkin, ``{R\'enyi entropy and conformal defects},'' {\em JHEP} {\bf 07} (2016) 076, \href{http://www.arXiv.org/abs/1511.06713}{{\tt 1511.06713}}.

\bibitem{Bhardwaj:2023kri}
L.~Bhardwaj, L.~E. Bottini, L.~Fraser-Taliente, L.~Gladden, D.~S.~W. Gould, A.~Platschorre, and H.~Tillim, ``{Lectures on generalized symmetries},'' {\em Phys. Rept.} {\bf 1051} (2024) 1--87, \href{http://www.arXiv.org/abs/2307.07547}{{\tt 2307.07547}}.

\bibitem{Shao:2023gho}
S.-H. Shao, ``{What's Done Cannot Be Undone: TASI Lectures on Non-Invertible Symmetries},'' \href{http://www.arXiv.org/abs/2308.00747}{{\tt 2308.00747}}.

\bibitem{Sasieta:2022ksu}
M.~Sasieta, ``{Wormholes from heavy operator statistics in AdS/CFT},'' {\em JHEP} {\bf 03} (2023) 158, \href{http://www.arXiv.org/abs/2211.11794}{{\tt 2211.11794}}.

\bibitem{Anous:2016kss}
T.~Anous, T.~Hartman, A.~Rovai, and J.~Sonner, ``{Black Hole Collapse in the 1/c Expansion},'' {\em JHEP} {\bf 07} (2016) 123, \href{http://www.arXiv.org/abs/1603.04856}{{\tt 1603.04856}}.

\bibitem{Balasubramanian:2022gmo}
V.~Balasubramanian, A.~Lawrence, J.~M. Magan, and M.~Sasieta, ``{Microscopic Origin of the Entropy of Black Holes in General Relativity},'' {\em Phys. Rev. X} {\bf 14} (2024), no.~1, 011024, \href{http://www.arXiv.org/abs/2212.02447}{{\tt 2212.02447}}.

\bibitem{Balasubramanian:2022lnw}
V.~Balasubramanian, A.~Lawrence, J.~M. Magan, and M.~Sasieta, ``{Microscopic Origin of the Entropy of Astrophysical Black Holes},'' {\em Phys. Rev. Lett.} {\bf 132} (2024), no.~14, 141501, \href{http://www.arXiv.org/abs/2212.08623}{{\tt 2212.08623}}.

\bibitem{Climent:2024trz}
A.~Climent, R.~Emparan, J.~M. Magan, M.~Sasieta, and A.~Vilar~L\'opez, ``{Universal construction of black hole microstates},'' {\em Phys. Rev. D} {\bf 109} (2024), no.~8, 086024, \href{http://www.arXiv.org/abs/2401.08775}{{\tt 2401.08775}}.

\bibitem{Geng:2024jmm}
H.~Geng and Y.~Jiang, ``{Microscopic Origin of the Entropy of Single-sided Black Holes},'' \href{http://www.arXiv.org/abs/2409.12219}{{\tt 2409.12219}}.

\bibitem{Balasubramanian:2024yxk}
V.~Balasubramanian, B.~Craps, J.~Hernandez, M.~Khramtsov, and M.~Knysh, ``{Factorization of the Hilbert space of eternal black holes in general relativity},'' {\em JHEP} {\bf 01} (2025) 046, \href{http://www.arXiv.org/abs/2410.00091}{{\tt 2410.00091}}.

\bibitem{Li:2024nft}
P.~Li, ``{Notes on the Factorisation of the Hilbert Space for Two-Sided Black Holes in Higher Dimensions},'' \href{http://www.arXiv.org/abs/2410.23886}{{\tt 2410.23886}}.

\bibitem{Hartman:2013mia}
T.~Hartman, ``{Entanglement Entropy at Large Central Charge},'' \href{http://www.arXiv.org/abs/1303.6955}{{\tt 1303.6955}}.

\bibitem{Asplund:2014coa}
C.~T. Asplund, A.~Bernamonti, F.~Galli, and T.~Hartman, ``{Holographic Entanglement Entropy from 2d CFT: Heavy States and Local Quenches},'' {\em JHEP} {\bf 02} (2015) 171, \href{http://www.arXiv.org/abs/1410.1392}{{\tt 1410.1392}}.

\bibitem{Lin:2016dxa}
F.-L. Lin, H.~Wang, and J.-j. Zhang, ``{Thermality and excited state R\'enyi entropy in two-dimensional CFT},'' {\em JHEP} {\bf 11} (2016) 116, \href{http://www.arXiv.org/abs/1610.01362}{{\tt 1610.01362}}.

\bibitem{Faulkner:2017hll}
T.~Faulkner and H.~Wang, ``{Probing beyond ETH at large $c$},'' {\em JHEP} {\bf 06} (2018) 123, \href{http://www.arXiv.org/abs/1712.03464}{{\tt 1712.03464}}.

\bibitem{Castro:2014tta}
A.~Castro, S.~Detournay, N.~Iqbal, and E.~Perlmutter, ``{Holographic entanglement entropy and gravitational anomalies},'' {\em JHEP} {\bf 07} (2014) 114, \href{http://www.arXiv.org/abs/1405.2792}{{\tt 1405.2792}}.

\bibitem{Chen:2023tpi}
B.~Chen, Y.~Liu, and B.~Yu, ``{Holographic complexity of rotating quantum black holes},'' {\em JHEP} {\bf 01} (2024) 055, \href{http://www.arXiv.org/abs/2307.15968}{{\tt 2307.15968}}.

\bibitem{Lauria:2018klo}
E.~Lauria, M.~Meineri, and E.~Trevisani, ``{Spinning operators and defects in conformal field theory},'' {\em JHEP} {\bf 08} (2019) 066, \href{http://www.arXiv.org/abs/1807.02522}{{\tt 1807.02522}}.

\bibitem{Kobayashi:2018okw}
N.~Kobayashi and T.~Nishioka, ``{Spinning conformal defects},'' {\em JHEP} {\bf 09} (2018) 134, \href{http://www.arXiv.org/abs/1805.05967}{{\tt 1805.05967}}.

\bibitem{Watanabe:2004nt}
T.~Watanabe and M.~J. Hayashi, ``{General relativity with torsion},'' \href{http://www.arXiv.org/abs/gr-qc/0409029}{{\tt gr-qc/0409029}}.

\bibitem{Giacomini:2006um}
A.~Giacomini, R.~Troncoso, and S.~Willison, ``{Junction conditions in general relativity with spin sources},'' {\em Phys. Rev. D} {\bf 73} (2006) 104014, \href{http://www.arXiv.org/abs/gr-qc/0603084}{{\tt gr-qc/0603084}}.

\bibitem{Brown:1986nw}
J.~D. Brown and M.~Henneaux, ``{Central Charges in the Canonical Realization of Asymptotic Symmetries: An Example from Three-Dimensional Gravity},'' {\em Commun. Math. Phys.} {\bf 104} (1986) 207--226.

\bibitem{Cardy:1986ie}
J.~L. Cardy, ``{Operator Content of Two-Dimensional Conformally Invariant Theories},'' {\em Nucl. Phys. B} {\bf 270} (1986) 186--204.

\bibitem{Zamolodchikov:1987avt}
A.~B. Zamolodchikov, ``{Conformal symmetry in two-dimensional space: Recursion representation of conformal block},'' {\em Theor. Math. Phys.} {\bf 73} (1987), no.~1, 1088--1093.

\bibitem{Hartman:2014oaa}
T.~Hartman, C.~A. Keller, and B.~Stoica, ``{Universal Spectrum of 2d Conformal Field Theory in the Large c Limit},'' {\em JHEP} {\bf 09} (2014) 118, \href{http://www.arXiv.org/abs/1405.5137}{{\tt 1405.5137}}.

\bibitem{Mathisson:1937zz}
M.~Mathisson, ``{Neue mechanik materieller systemes},'' {\em Acta Phys. Polon.} {\bf 6} (1937) 163--200.

\bibitem{Papapetrou:1951pa}
A.~Papapetrou, ``{Spinning test particles in general relativity. 1.},'' {\em Proc. Roy. Soc. Lond. A} {\bf 209} (1951) 248--258.

\bibitem{Dixon:1970zza}
W.~G. Dixon, ``{Dynamics of extended bodies in general relativity. I. Momentum and angular momentum},'' {\em Proc. Roy. Soc. Lond. A} {\bf 314} (1970) 499--527.

\bibitem{Fitzpatrick:2016ive}
A.~L. Fitzpatrick, J.~Kaplan, D.~Li, and J.~Wang, ``{On information loss in AdS$_{3}$/CFT$_{2}$},'' {\em JHEP} {\bf 05} (2016) 109, \href{http://www.arXiv.org/abs/1603.08925}{{\tt 1603.08925}}.

\bibitem{Fitzpatrick:2016mjq}
A.~L. Fitzpatrick and J.~Kaplan, ``{On the Late-Time Behavior of Virasoro Blocks and a Classification of Semiclassical Saddles},'' {\em JHEP} {\bf 04} (2017) 072, \href{http://www.arXiv.org/abs/1609.07153}{{\tt 1609.07153}}.

\bibitem{Chen:2017yze}
H.~Chen, C.~Hussong, J.~Kaplan, and D.~Li, ``{A Numerical Approach to Virasoro Blocks and the Information Paradox},'' {\em JHEP} {\bf 09} (2017) 102, \href{http://www.arXiv.org/abs/1703.09727}{{\tt 1703.09727}}.

\bibitem{Skenderis:2008dh}
K.~Skenderis and B.~C. van Rees, ``{Real-time gauge/gravity duality},'' {\em Phys. Rev. Lett.} {\bf 101} (2008) 081601, \href{http://www.arXiv.org/abs/0805.0150}{{\tt 0805.0150}}.

\bibitem{Skenderis:2008dg}
K.~Skenderis and B.~C. van Rees, ``{Real-time gauge/gravity duality: Prescription, Renormalization and Examples},'' {\em JHEP} {\bf 05} (2009) 085, \href{http://www.arXiv.org/abs/0812.2909}{{\tt 0812.2909}}.

\bibitem{Dolan:2000ut}
F.~A. Dolan and H.~Osborn, ``{Conformal four point functions and the operator product expansion},'' {\em Nucl. Phys. B} {\bf 599} (2001) 459--496, \href{http://www.arXiv.org/abs/hep-th/0011040}{{\tt hep-th/0011040}}.

\bibitem{Larkin:1964wok}
A.~I. Larkin and Y.~N. Ovchinnikov, ``{Nonuniform state of superconductors},'' {\em Zh. Eksp. Teor. Fiz.} {\bf 47} (1964) 1136--1146.

\bibitem{Roberts:2014ifa}
D.~A. Roberts and D.~Stanford, ``{Two-dimensional conformal field theory and the butterfly effect},'' {\em Phys. Rev. Lett.} {\bf 115} (2015), no.~13, 131603, \href{http://www.arXiv.org/abs/1412.5123}{{\tt 1412.5123}}.

\bibitem{Rozenbaum:2016mmv}
E.~B. Rozenbaum, S.~Ganeshan, and V.~Galitski, ``{Lyapunov Exponent and Out-of-Time-Ordered Correlator\textquoteright{}s Growth Rate in a Chaotic System},'' {\em Phys. Rev. Lett.} {\bf 118} (2017), no.~8, 086801, \href{http://www.arXiv.org/abs/1609.01707}{{\tt 1609.01707}}.

\bibitem{Shenker:2013pqa}
S.~H. Shenker and D.~Stanford, ``{Black holes and the butterfly effect},'' {\em JHEP} {\bf 03} (2014) 067, \href{http://www.arXiv.org/abs/1306.0622}{{\tt 1306.0622}}.

\bibitem{Shenker:2013yza}
S.~H. Shenker and D.~Stanford, ``{Multiple Shocks},'' {\em JHEP} {\bf 12} (2014) 046, \href{http://www.arXiv.org/abs/1312.3296}{{\tt 1312.3296}}.

\bibitem{Shenker:2014cwa}
S.~H. Shenker and D.~Stanford, ``{Stringy effects in scrambling},'' {\em JHEP} {\bf 05} (2015) 132, \href{http://www.arXiv.org/abs/1412.6087}{{\tt 1412.6087}}.

\bibitem{Kobayashi:2018lil}
N.~Kobayashi, T.~Nishioka, Y.~Sato, and K.~Watanabe, ``{Towards a $C$-theorem in defect CFT},'' {\em JHEP} {\bf 01} (2019) 039, \href{http://www.arXiv.org/abs/1810.06995}{{\tt 1810.06995}}.

\bibitem{Cuomo:2021rkm}
G.~Cuomo, Z.~Komargodski, and A.~Raviv-Moshe, ``{Renormalization Group Flows on Line Defects},'' {\em Phys. Rev. Lett.} {\bf 128} (2022), no.~2, 021603, \href{http://www.arXiv.org/abs/2108.01117}{{\tt 2108.01117}}.

\bibitem{Popov:2025cha}
F.~K. Popov and Y.~Wang, ``{Factorizing Defects from Generalized Pinning Fields},'' \href{http://www.arXiv.org/abs/2504.06203}{{\tt 2504.06203}}.

\end{thebibliography}\endgroup

\end{document}